\documentclass[12pt,preprint]{aastex}

\newcommand{\Msun}{\mbox{$M_{\odot}$}}

\begin{document}

\title{The Structure of the Accretion Disk in the Accretion Disk Corona X-Ray
Binary 4U 1822--371 at Optical and Ultraviolet Wavelengths}

\author{Amanda J. Bayless} \author{Edward L. Robinson}
\affil{Department of Astronomy, University of Texas at Austin\\ 1
University Station, Austin, TX 78712-0259, USA}
\author{Robert I. Hynes} {} \affil{Department
of Physics and Astronomy, Louisiana State University,\\ Baton Rouge,
LA 70803, USA}
\author{Teresa A. Ashcraft} \affil{School of Earth and Space Exploration, Arizona State University,
Tempe, AZ 85287-1404, USA}
\and \author{Mark E. Cornell} \affil{Department of Astronomy,
University of Texas at Austin\\ 1 University Station, Austin, TX
78712-0259, USA}

\begin{abstract}
The eclipsing low-mass X-ray binary 4U 1822--371 
is the prototypical accretion disk corona (ADC) system.   
We have obtained new time-resolved UV spectroscopy of
4U~1822--371 with the Advanced Camera for Surveys/Solar Blind Channel on the {\it Hubble Space Telescope} and new $V$- and $J$-band 
photometry with the 1.3-m SMARTS telescope at Cerro Tololo Inter-American Observatory. 
We use the new data to construct its UV/optical spectral energy 
distribution and its orbital light curve in the UV, $V$, 
and $J$ bands. 
We derive an improved ephemeris for the optical
eclipses and confirm that the orbital period is changing 
rapidly, indicating extremely high rates of mass flow in
the system; and we show that the accretion disk in the system
has a strong wind with projected velocities up to 4000~km~s$^{-1}$.
We show that the disk has a vertically-extended, 
optically-thick component at optical wavelengths.
This component extends almost to the
edge of the disk and has a height equal to $\sim$0.5 
of the disk radius.
As it has a low brightness temperature, we identify it
as the optically-thick base of
a disk wind, not as the optical counterpart of the ADC.
Like previous models of 4U~1822--371, ours needs
a tall obscuring wall near the edge of the
accretion disk, but we interpret the wall as a layer of cooler
material at the base of the disk wind, not as a tall, luminous 
disk rim.
\end{abstract}


\section{Introduction}
4U~1822--371 is a low-mass X-ray binary star (LXMB) with an
orbital period of $P_{orb} = 5.57$~hr \citep{mas80}. 
The compact star in the system is an X-ray pulsar with spin period
$P_{pulse} = 0.593$~s \citep{jon01}  and is, therefore, a neutron star.
The UV/optical spectrum of 4U~1822--371 is rich with
emission and absorption lines 
\citep{cha80, mas82a, mas82c, har97, cow82, cow03, jon03, hut05}.
Doppler tomography of the \ion{He}{2}~$\lambda4686$
and \ion{O}{6}~$\lambda3811$ emission lines confirms
expectations that there is an accretion disk
around the neutron star fed by a stream of gas
from the secondary star \citep{cas03}.

4U~1822--371 is one of the rare eclipsing LXMBs \citep{mas80, whi81}.
Many analyses of the eclipse have been published
\citep{whi81, whi82, mas82a, mas82b, hel89, bap02, cow03}
and, although the detailed results of the various analyses often
differ greatly, there is substantial agreement on several points.
The eclipse is
a transit of the secondary star across the accretion disk and, since
the eclipse is very broad at optical wavelengths, the accretion disk is large,
although exactly how large is uncertain:
\citet{mas82b} find $r_{\rm disk}/a = 0.28$ to 0.38 and
\citet{hel89} find $r_{\rm disk} / a = 0.58\pm 0.08$, where 
$r_{\rm disk}$ is the radius of the disk and $a$ is the
separation of the stars.
The X-ray eclipse is also broad. Its flux at minimum is at about 50\% of the uneclipsed flux level
and it has no sharp features, showing that the X-ray emission comes from 
a partially-eclipsed extended cloud around the neutron star -
an accretion disk corona (ADC) \citep{whi81,mas82a}.
The corona extends to $\sim 0.5 r_{\rm disk}$ in the plane of
the disk and is vertically extended. 
Estimates of the ratio of its height to its radius 
in the plane of the disk range from 0.3 to 1 
\citep{whi82, mas82a, hel89}.
The orbital inclination lies in the range
$78^\circ < i < 84^\circ$.

The ratio of the X-ray luminosity of 4U1822--371 to its 
optical luminosity, $L_X/L_{opt}$, lies between 
$\sim 15$ and $\sim 65$, 
much less than that for typical LXMBs, indicating that 
the ADC is optically thick and blocks most of the 
X-ray emission from the inner disk and neutron star \citep{gri78,mas80}.
Fits to the X-ray spectrum using realistic 
Comptonization models yield electron temperatures near 
$T_e \approx 5 \times 10^7$~K and electron scattering 
optical depths in the range $\tau_{e}=13$--26 depending 
on the geometry of the ADC and the Comptonization model
\citep{par00, iar01}.
The mean unabsorbed apparent luminosity of 4U~1822--371 
in the 0.1--100 keV energy range is 
$1.15 \times 10^{36}\ \textrm{erg s}^{-1}$ for a distance
of 2.5~kpc \citep{iar01}.
As most of the X-ray flux escapes roughly perpendicularly
to the orbital plane and is beamed away from the Earth, the true 
luminosity is much greater than the apparent luminosity.
If one corrects for the ADC obscuration by
simply multiplying $L_X$ by a factor to
raise $L_X/L_{opt}$ to the typical value for LXMBs 
($\sim 500$), the true X-ray luminosity is
$\sim 10^{37}\ \textrm{erg s}^{-1}$.
Even this should be taken as a lower limit because 
the disk is observed edge-on, reducing $L_{opt}$.
A true luminosity near the Eddington luminosity
is not impossible.

All solutions for the X-ray eclipse have invoked a vertically-extended
 wall of optically-thick material whose height varies with 
angle around the neutron star \citep{whi81}.
The wall generally has been identified as a vertically extended
disk rim, and, indeed,
\citet{hel89} showed the wall must be near the edge of
the disk.  The rim invoked by \citet{hel89} 
has $0.08 < h_{rim}/r_{\rm disk} < 0.22$, while the maximum height of
the rim invoked by \citet{whi82} approached
$h_{rim}/r_{\rm disk} \approx 0.5$.
\citet{iar01} also needed a thick belt of absorbing material 
around the ADC to match the observed X-ray spectral energy distribution (SED).
Placing the belt at the edge of the disk, they found the 
angle subtended by the absorbing
region is $16^\circ$, or $h_{rim}/r_{\rm disk} \approx 0.29$.
These rims are an order of magnitude taller than the
expected thickness of the disk proper and cannot
be pressure supported.
\citet{whi82} invoked turbulence to support the rim
but the turbulence would have to be highly supersonic,
needing Mach numbers between 7 and 15.
While it is not inconceivable that supersonic turbulence 
could be generated by the impact of the accretion stream on
the disk, it is unclear how supersonic turbulence could be 
maintained around the entire circumference of the disk.

It has been universally assumed that the only contributors to
the UV/optical continuum are 1) the accretion disk, 2) its extended
rim, which may be irradiated and have a higher temperature on
its surface towards the neutron star, and 3) the irradiated secondary
star \citep{mas82b,hel89,puc95}.
It also has been assumed that the ADC is optically thin at 
UV/optical wavelengths, even though it is optically thick to
electron scattering at X-ray wavelengths. These contradictory preclude a UV/optical contribution from the ADC or from vertical structures other than the rim.

The ADC and the eclipses lend particular importance to 4U~1822--371.
ADC sources are more numerous than suggested by the
small number that have been identified.
The luminosity of an ADC is typically $\sim100$ times 
less than the luminosity of the inner disk and neutron star,
so ADCs are visible only in those systems with orbital inclinations
near enough to $90^\circ$ that the outer accretion disk or the 
ADC itself blocks most of the flux from the more luminous
components of the binary \citep{whi82}.
Many ADC sources must be lurking among the low-inclination 
LXMBs, perhaps 5 -- 10 times more than are currently known.
4U~1822--371 is the prototype for these systems.
Models for accretion onto neutron stars and black
holes in X-ray binaries or onto supermassive black holes in
AGN need much more than flat, unadorned accretion disks.
They must also include high-temperature flows, vertical extension,
winds, and irradiation, all of which can vary with time
\citep{nar98,bla99,wij99,nar00,las01}.
ADCs are examples of high-temperature accretion flows
and vertically-extended structures.
The eclipses of 4U~1822--371 offer the opportunity to
map these structures in detail.

We have obtained new {\it HST} objective-prism ultraviolet
spectroscopy and $V$ and $J$ photometry of 4U~1822--371, from
which we construct its SED and its orbital light curves in the UV, $V$, and $J$ bands
(Section 2).
In Section 3 we derive an improved ephemeris for the optical
eclipses and confirm that the orbital period is changing 
rapidly, indicating extremely high rates of mass flow in
the system.
In Section 4 we show that the accretion disk 
has a strong wind with projected
velocities up to 4000~km~s$^{-1}$.
The core of this paper is Section 5, in which we model the 
orbital light curves with our light curve synthesis program. 
We show that much of the disk is vertically extended and optically
thick at UV/optical wavelengths.
This vertical structure extends almost to the
edge of the disk and has a height equal to $\sim$0.5 of the disk radius.
As this structure has a brightness temperature near
$3 \times 10^4$~K, we identify it as the optically-thick base of
the disk wind, \emph{not} as the optical counterpart of the ADC.
We, too, need a tall obscuring wall near the edge of the
accretion disk, but we interpret the wall as layer of cooler
material at the base of the disk wind, not as a tall
disk rim.
Our results are summarized in Section 6.

\section{Observations and Data Reduction}
We observed 4U 1822--371 with the ANDICAM optical/IR photometer on the
SMARTS 1.3-m telescope at CTIO using Johnson $V$ and CTIO/CIT $J$
filters \citep{andicam}. 
We obtained one or two observations per night over 
two observing seasons, the first season from 2005
June 30 to September 15, the second from 2006 March 29 to October 20. 
The data were reduced to instrumental magnitudes with 
IRAF\footnote{IRAF is distributed by the
National Optical Astronomy Observatory, which is operated by the
Association of Universities for Research in Astronomy, Inc., under the
cooperative agreement with the National Science Foundation},
and the
instrumental magnitudes were converted to absolute fluxes using the
photometric standard stars PG1657+078 for the $V$ band and 2MASS
J18254777-3706131 for the $J$ band \citep{lan92}, both of
which were observed on three photometric nights in 2007 June.

We also obtained a time series of objective-prism 
ultraviolet spectrograms of 4U 1822--371 with the 
{\it Hubble Space Telescope (HST)}
using the Advanced Camera for Surveys/Solar Blind Channel (ACS/SBC) and the PR130L prism \citep{acs}.
We obtained 300 spectrograms in two visits between 
HJD~2453828.7 and HJD~2453830.9 (2006 April), 
each visit lasting five {\it HST} orbits.  
The spectrograms were integrated for 45 seconds and
separated by a 40-second dead time, with occasional 
long exposures of 350 seconds when data were transfered 
from the ACS internal buffer
memory to the {\it HST} solid state data recorder.  
The spectrograms cover
the wavelength range from 1222~\AA\ to 2002~\AA,
but we used only the wavelengths from 1222 to 1900~\AA\ 
because there is a significant red leak in
the SBC at wavelengths greater than 1900~\AA.  
The spectral resolution is $R=\lambda/\Delta\lambda\approx150$ 
near 1300~\AA, but decreases to $R\approx100$ near 
the \ion{C}{4} doublet and to $R\approx40$ near 1900 \AA.  
The UV data were reduced
with the {\it aXe} package in STSDAS \citep{axe}.
  
\begin{figure}
\center \includegraphics[angle=270,scale=.65]{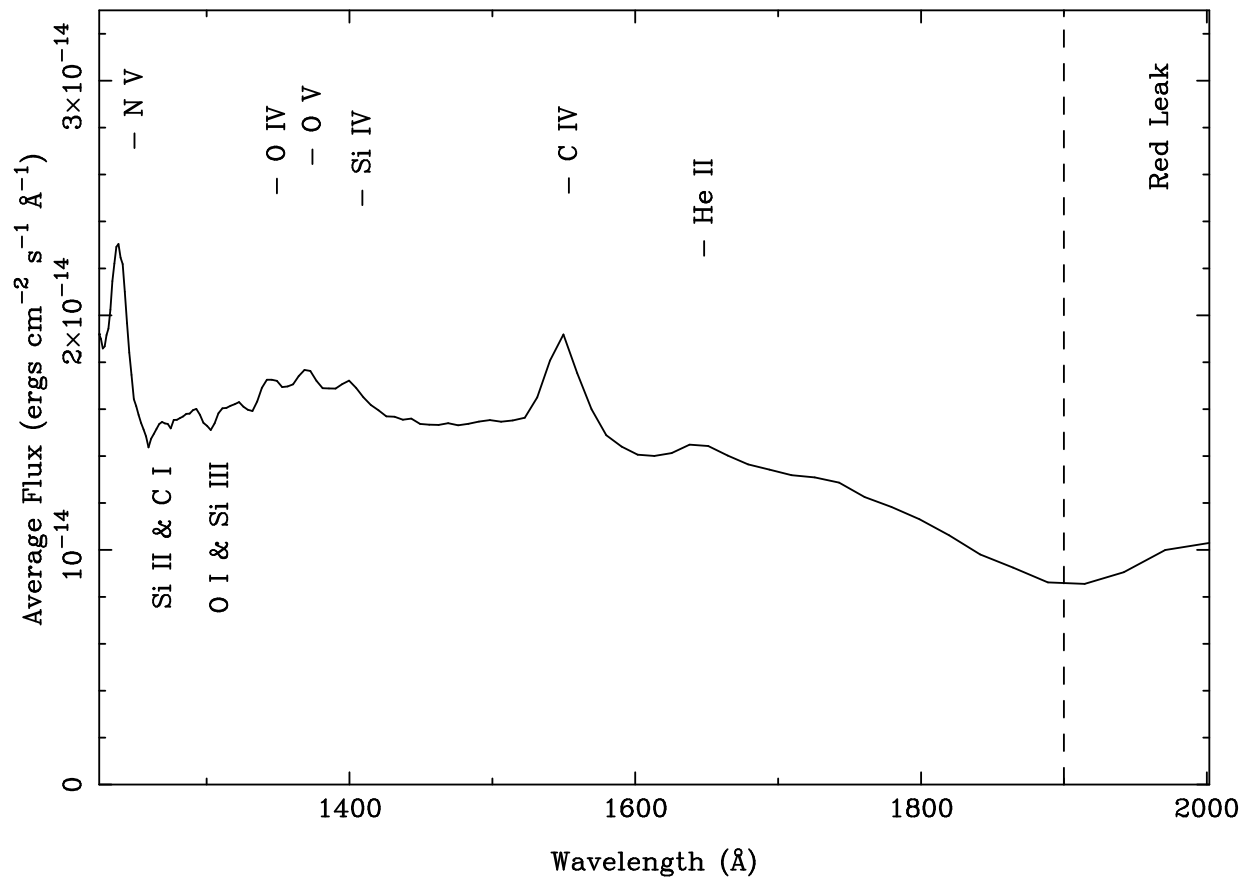}
\caption{The UV spectrum of 4U 1822--371.  
The prominent
emission lines are \ion{N}{5} at 1240~\AA, a blend of \ion{O}{4},
\ion{O}{5}, and \ion{Si}{4} near 1370~\AA, the \ion{C}{4} doublet at 
1548~\AA/1550 \AA, and \ion{He}{2} at~1640 \AA.  
There are also ISM
absorption blends of  \ion{Si}{2}/\ion{C}{1} and
\ion{O}{1}/\ion{Si}{3} near 1260~\AA\ and 1300~\AA\ respectively.
There is a significant red leak in the SBC at wavelengths longer
than $\sim1900$~\AA.}
\label{specUV}
\end{figure}

The time-averaged mean UV spectrum of 4U 1822--371
is shown in Figure \ref{specUV}.  
The prominent
emission lines are \ion{N}{5} at 1240~\AA, a blend of \ion{O}{4},
\ion{O}{5}, and \ion{Si}{4} near 1370~\AA, the \ion{C}{4} doublet at 
1548/1550~\AA, and \ion{He}{2} at~1640 \AA.  
There are also interstellar absorption lines from
blends of \ion{Si}{2}/\ion{C}{1} near 1260~\AA\ and
\ion{O}{1}/\ion{Si}{3} near 1300~\AA.

\section{The UV/Optical Ephemeris and Orbital Light Curves}

Because the $V$ and $J$ light curves are sparsely sampled, it was not
possible to extract times of minima for individual eclipses.  
Instead we measured a single mean time of minimum for each 
season by folding the $V$-band data on the orbital period 
from \citet{par00} to form seasonal mean orbital light curves.   
The times of minimum light extracted from the
mean light curves are $T_{\rm min} = $ HJD $ 2453618.677 \pm 0.005$ for the
2005 data and $T_{\rm min} = $ HJD $2453932.720 \pm 0.004$ for the 2006
data.
To form UV light curves the flux in each UV spectrogram was integrated
from 1222~\AA\ to 1900~\AA\ and then divided by 
$\Delta \lambda = 678$~\AA.
We folded the light curves on the
\citet{par00} period and extracted two times of minimum light,
one for each {\it HST} visit: 
$T_{\rm min} = \textrm{HJD}\ 2453828.972\pm0.001$ and 
$T_{\rm min} = \textrm{HJD}\ 2453830.830\pm0.001$.

\begin{figure}
\center \includegraphics[angle=270,scale=.65]{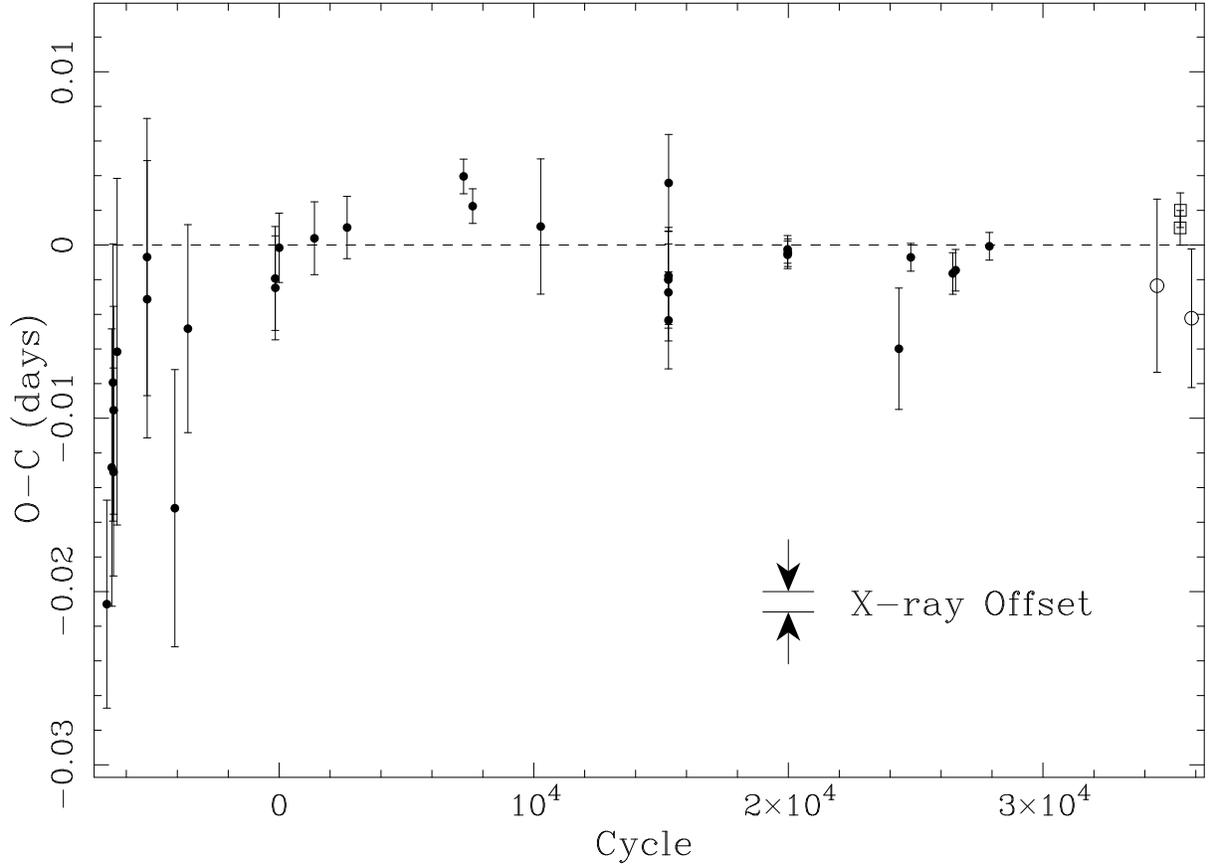}
\caption{The $O-C$ diagram for the optical/UV eclipse
  ephemeris (equation~\ref{Ephemeris}).
  The open circles are the $V$ eclipse times from SMARTS data and 
  the open squares are the UV eclipse times from {\it HST} data. 
  The solid circles are the previously published times of optical
  eclipses listed in Table \ref{time}.  
  The arrows show the mean offset between the ephemerides for
  the X-ray and UV/optical eclipses.
\label{OCfitopt}}
\end{figure}

The four new eclipse times plus all of the previously published 
optical eclipse times are listed in Table~\ref{time}.   
A fit to the eclipse times yielded the ephemeris
\begin{equation}
T_{\rm min} = \textrm{HJD}\ 2445615.31166(74) + 0.232108641(80)E
+2.46(21)\times10^{-11}E^2, \label{Ephemeris}
\end{equation}
where  $E$ is the eclipse number.  
The $O-C$ diagram for this ephemeris is
shown in Figure \ref{OCfitopt}.  
The quadratic term in the optical ephemeris is consistent
with the low-significance quadratic term derived by 
\citet{bap02} and agrees with
the quadratic term in the X-ray ephemeris,
$(2.06\pm23)\times10^{-11}$, to within the measurement errors \citep{par00}. 
The $T_{\rm min}$ of our new optical ephemeris lags the $T_{\rm min}$ from the X-ray ephemeris of \citet{par00} by $100 \pm 65$ seconds. 
Although the lag is only marginally statistically significant, 
it cannot be dismissed because the X-ray and 
optical/UV flux emanate from different regions and 
could be eclipsed at slightly different times.  Recently, Burderi et al.  (private communication) derived a new X-ray ephemeris for 4U~1822-371.  Using this new ephemeris the lag is 122 seconds.

The rate of change of the orbital period is
${\dot P}=2.12\times10^{-10}$ and the timescale for
a change in orbital period is
${{P} / {\dot P}} = (3.0 \pm 0.3) \times 10^{6}\ \textrm{yr}$.
One must generally be cautious about interpreting
timescales for orbital period changes
as evolutionary timescales because
mass-transfer binaries often
show short-term systematic advances and delays in times 
of eclipse that can mimic rapid period changes.
The X-ray binary EXO 0748-676 shows this behavior \citep{wol09} and there are many examples among cataclysmic variables 
\citep[see][]{war95}. 
However, the eclipse times for 4U~1822--371 span 27 years 
and the accumulated phase shift is 
$\Delta \phi = \Delta T_{\rm min}/P \approx 0.17$, an
order of magnitude larger than the random phase
shifts observed in EXO 0748-676 and cataclysmic variables. 
It appears safe, therefore, to identify the value of
$P/\dot P$ measured from the ephemeris as the time 
scale for orbital evolution.

On dimensional grounds one has
${{m} / {\dot m}} \approx \vert P/\dot P\vert$.  If mass and angular
momentum are conserved, for example, then $m_{NS}/\dot m_{NS} = [3(1-q)/q] P/\dot P$ \citep{war95},
where $m_{NS}$ is the mass of the neutron star, $q=m_2/m_{NS}$, and $m_2$ is the mass of the secondary star.
For $q=0.3$ and $m_{NS}=1.35$\Msun, we find $\dot m_{NS}=6.4\times10^{-8}\Msun$ yr$^{-1}$.
This is an extremely high rate of accretion, corresponding to $\sim 4$ times the Eddington luminosity.
In this we agree with the analysis of \citet{hei01}.
We do not agree that the flow of mass and angular momentum
can be characterized by a single parameter.
Specifically, the angular momentum per unit mass can vary 
enormously in various parts of the flow and depends 
on the detailed properties of the flow.
Absent a precise description for the flow of mass and
angular momentum, it is not possible to infer accurate 
rates of mass transfer and mass loss from the 
evolutionary timescale, nor even a ratio of the two
rates.
Few cataclysmic variables have rates of mass transfer
greater than $3 \times 10^{-8}\ M_\odot \ \textrm{yr}^{-1}$
\citep{war95},
so we suspect that the rate of mass transfer in 
4U~1822--371 is at least a factor of two smaller than the value for conservative mass transfer.
Yet, a mass transfer rate of $3 \times 10^{-8}\ M_\odot \ \textrm{yr}^{-1}$ is still
extremely high and much of the transferred mass cannot 
be accreting onto the neutron star, but leaves the system altogether.

\begin{figure}
\center \includegraphics[angle=270,scale=.65]{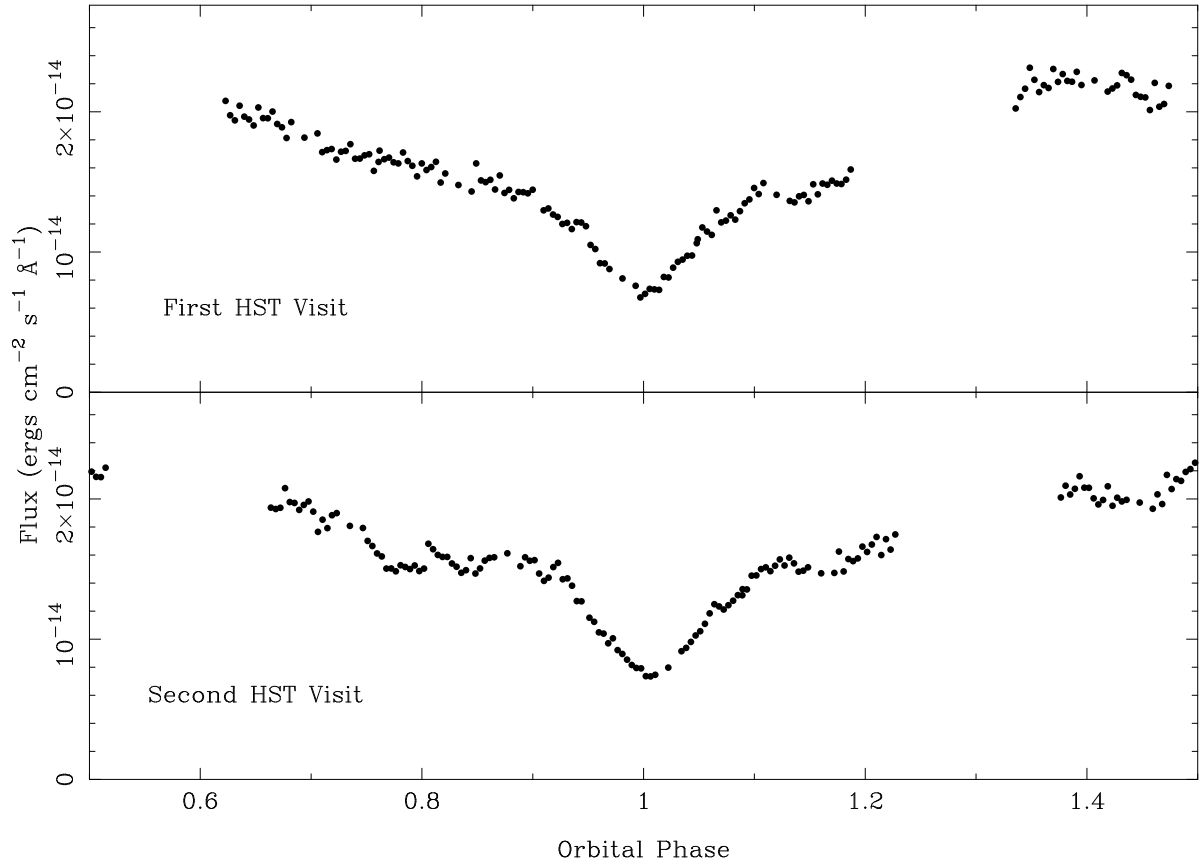}
\caption{The UV light curve of 4U~1822--371 from 2006.
  The fluxes were extracted from the
  {\it HST} spectrograms and folded on the orbital period of 4U~1822--371 (equation~\ref{Ephemeris})  to produce one light curve for each of the {\it HST} visits.
  The internal errors on the fluxes are $\sim 0.5$\%, giving
  error bars smaller than the points used to plot the data.}
\label{uvlc}
\end{figure}

\begin{figure}
\center \includegraphics[angle=270,scale=.65]{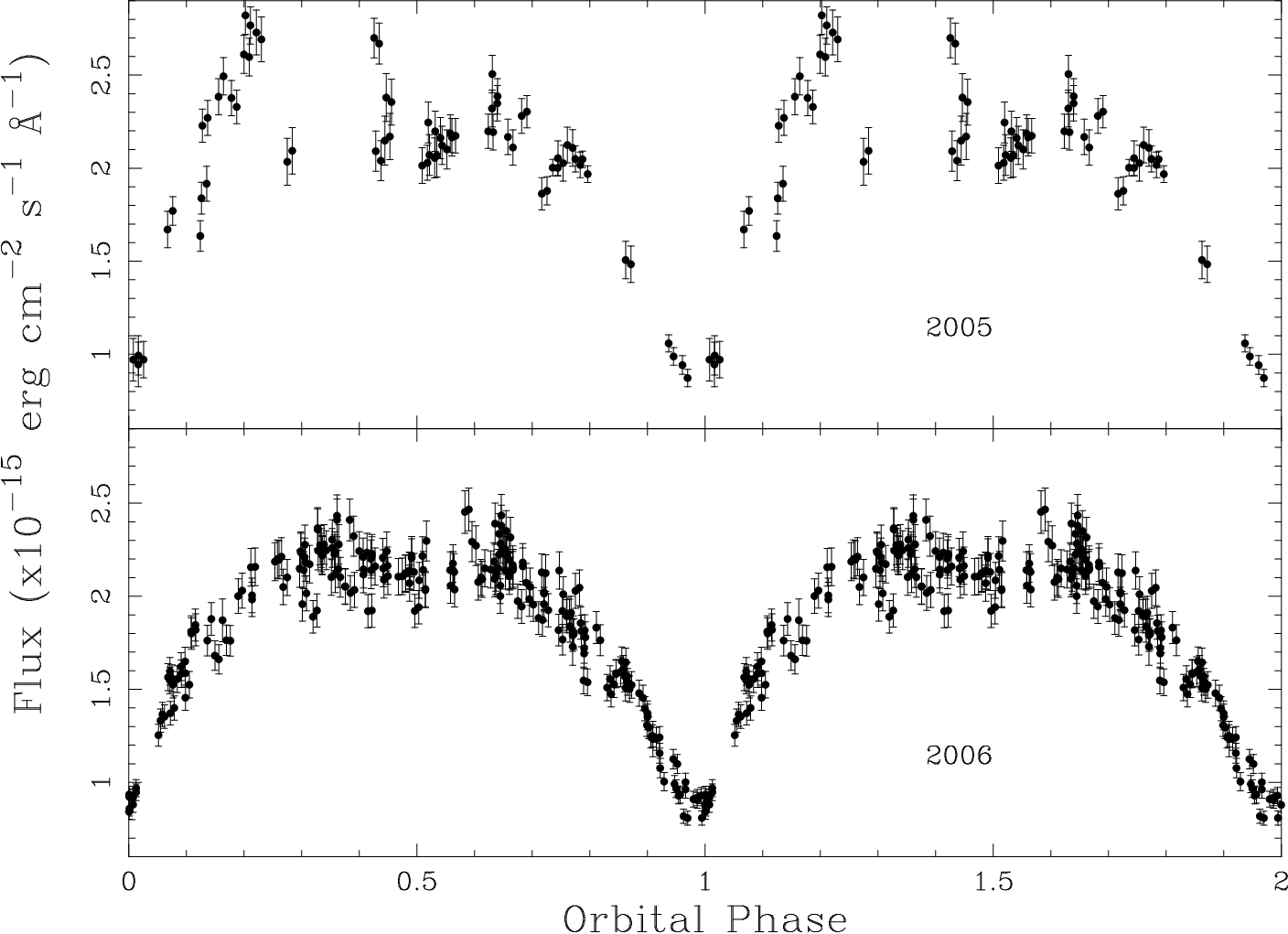}
\caption{$V$-band light curves of 4U~1822--371 in 2005 and 2006.
  The data have been folded on the orbital period given by
  the UV/optical ephemeris (equation~\ref{Ephemeris}).
  The error bars correspond to the internal measurement errors.}
\label{LCv}
\end{figure}

\begin{figure}
\center \includegraphics[angle=270,scale=.65]{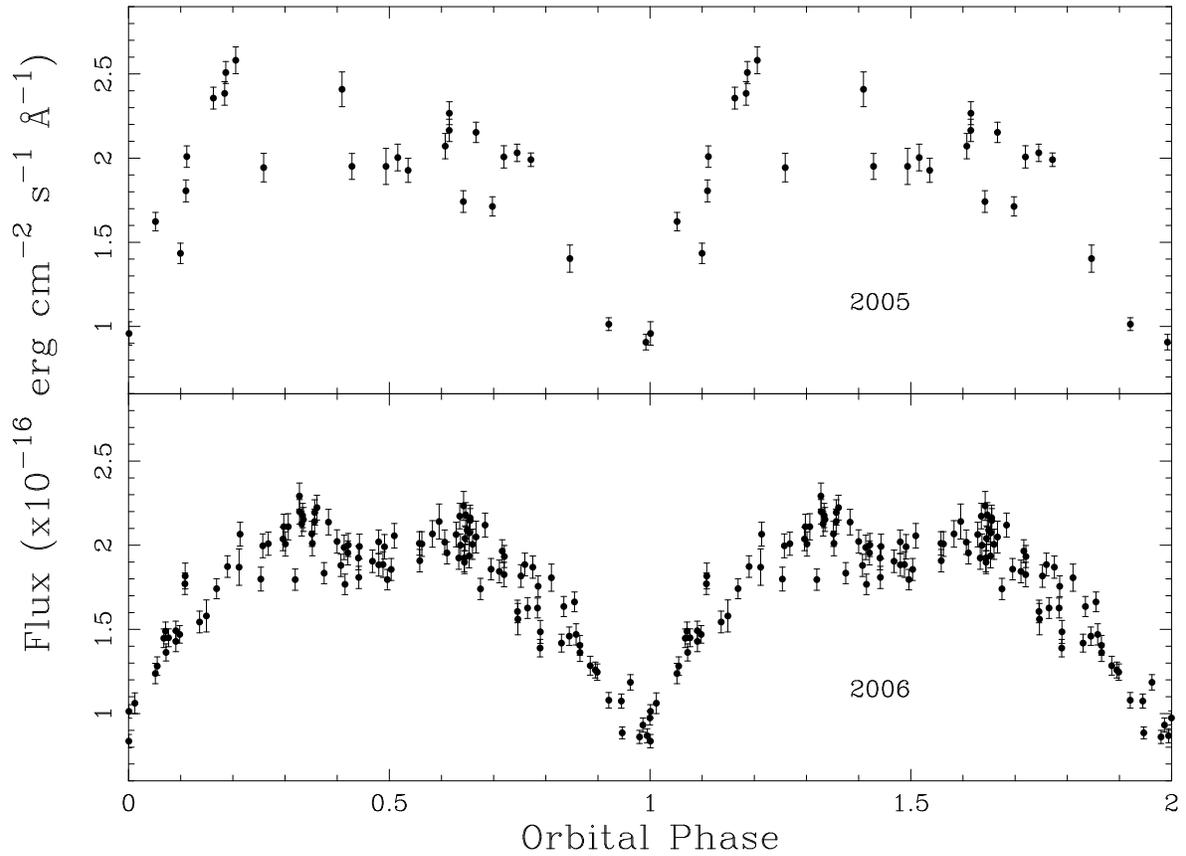}
\caption{$J$-band light curves of 4U~1822--371 in 2005 and 2006.
  The data have been folded on the orbital period given by
  the UV/optical ephemeris (equation~\ref{Ephemeris}).
  The error bars correspond to the internal measurement errors.}
\label{LCj}
\end{figure}

Figure~\ref{uvlc} shows the UV light curves of 4U~1822--371
from the two {\it HST} visits folded on the 
orbital period given by equation~\ref{Ephemeris}; 
and Figures~\ref{LCv} and
\ref{LCj} show the two seasons of $V$- and $J$-band
photometry also folded on the orbital period. 
The folded UV light curves and the $V$-band light curve from
2006 have been averaged into 200 equal-width
phase bins.
The $J$-band light curves from both 2005 and 2006
and the $V$-band light curve from 2005 have fewer observations,
and have not been binned.  We define the internal error of a data point
to be the standard deviation that includes
photon counting noise and calibration errors.
The internal errors of the UV fluxes are small, 
typically only 0.5\%;
the internal errors of $V$- and $J$-band data were
typically 4 -- 5\% but occasionally somewhat larger.
The error bars in the figures correspond to these internal
errors.

The light curve of 4U~1822--371 shows large-amplitude
non-repeating variability (``flickering'') and, because
of this variability, individual orbital light
curves can differ substantially from the mean orbital light 
curve \citep{bap02}.
Let us define the external error of a data point 
as the standard deviation
introduced by the non-repeating variability.
We estimated the external error of the UV fluxes by 
comparing the light curves
of 4U~1822--371 from the two {\it HST} visits. 
After folding all the UV data together to form a single
(noisy) light curve, we divided the light curve into four 
phase sections 
(0.11 -- 0.23, 0.33 -- 0.52, 0.62 -- 0.9, and 0.9 -- 1.11)
and fit a polynomial to each section.
The standard deviation of the data points about the fitted
polynomial ranged from 5\% to 9\%.
We equate the external error of the data points in each section
to the standard deviation for the section.  
Much the same method was used to determine the external error in 
the $V$- and $J$-band data except that
the light curves were divided into just
two sections covering phases 0.25 -- 0.70 and phases 0.70 -- 1.25.
The error bars in Figure \ref{3lc} correspond to these 
external errors.
When calculating the $\chi^2$ of models fitted to the
light curves (Section 5), we will use the larger of the internal and
external errors for each data point.
With the exception of a few data points in the 
$V$- and $J$-bands, the external errors are much larger
than the internal errors and it is these large external
errors that dominate the fit.

The shape of the $V$- and $J$-band light curves changed 
systematically between 2005 and 2006.
While the 2006 light curves are roughly symmetrical about phase 0.0, 
the 2005 light curves are asymmetric, having a large hump near phase 0.2.
The asymmetric light curves from 2005 are similar to the 
optical light curves shown in
\citet{mas80}, \citet{har97}, and \citet{bap02}.
The more symmetric light curves from 2006 are similar to our UV light 
curves -- also obtained in 2006 -- and to the optical light curves
shown in \citet{cow03}.

\begin{figure}
\center \includegraphics[angle=270,scale=.65]{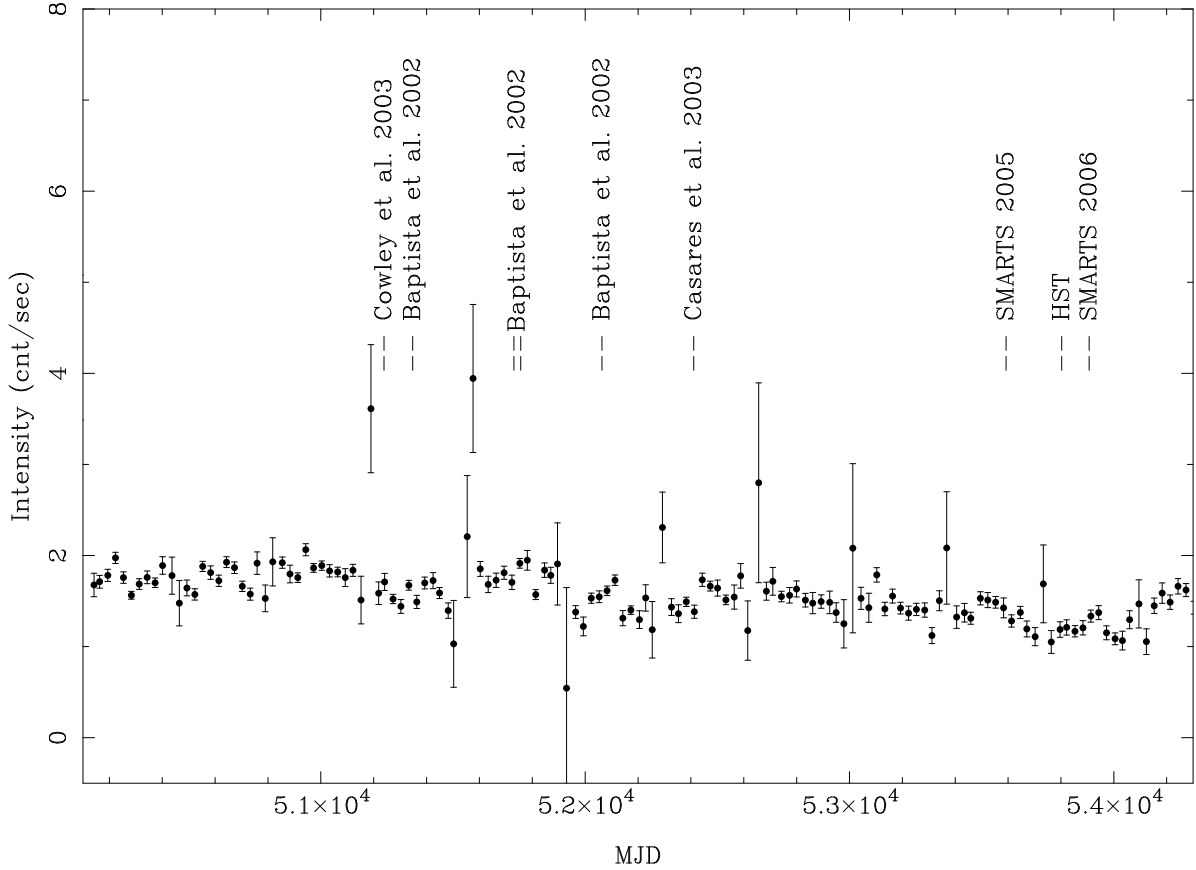}
\caption{The X-ray light curve of 4U 1822--371 
  between 1996 January 5 and 2008 March 5  
  from the All Sky Monitor on RXTE.
  Each point
  is the monthly average of all three ASM bandpasses.
  The vertical dashes mark the times of eclipse minima 
  at UV/optical wavelengths measured by
  \citet{cow03}, \citet{bap02} (4 eclipses), \citet{cas03}, 
  and by us with {\it HST} and SMARTS.}
\label{xte}
\end{figure}

Figure \ref{xte} shows the X-ray light curve of 4U~1822--371
between 1996 January 5 and 2008 March 5 from the
All Sky Monitor (ASM) on the {\it Rossi
X-ray Timing Explorer} (RXTE)\footnote{Results provided by
the ASM/RXTE teams at MIT and at the RXTE SOF and GOF at NASA's GSFC.
http://xte.mit.edu}. 
Each point in the figure is the monthly average count 
rate summed over the three ASM passbands.
The dates on which optical eclipses were observed are marked 
on the figure.  
The X-ray light curve shows a low-amplitude
variability superimposed on a slow decrease in flux.
The pronounced changes in the optical/UV light curves 
from year to year were not accompanied by large changes 
in the X-ray light curve. 

\section{The UV Spectrum, UV/Optical 
Spectral Energy Distributions, and Evidence for a Strong
High-Velocity Wind}

Although the direction to 4U~1822--371 is within $11^\circ$
of the direction to the galactic center, 4U~1822--371 is
close to Baade's window and its reddening is low.
\citet{mas82a} find a color excess $E(B-V) = 0.1$ with 
$3\sigma$ upper and lower bounds of 0.29 and 0.01 based on the
depth of the 2200~\AA\ interstellar absorption feature.
According to \citet{iar01} the photoelectric absorption 
at soft X-ray energies corresponds to a neutral hydrogen column
density of $N_H \approx (1.05\pm 0.45) \times 10^{21}$~cm$^{-2}$.
Adopting $A_V = N_H/(2.21\pm0.09) \times 10^{21}$~mag~cm$^2$ \citep{guv09}
and $R_V=3.1$, we find 
$A_V = 0.48 \pm 0.2$ mag and 
$E(B-V) = 0.15 \pm 0.1$ mag in agreement with \citet{mas82a}.

\begin{figure}
\center \includegraphics[scale=0.65, angle=0]{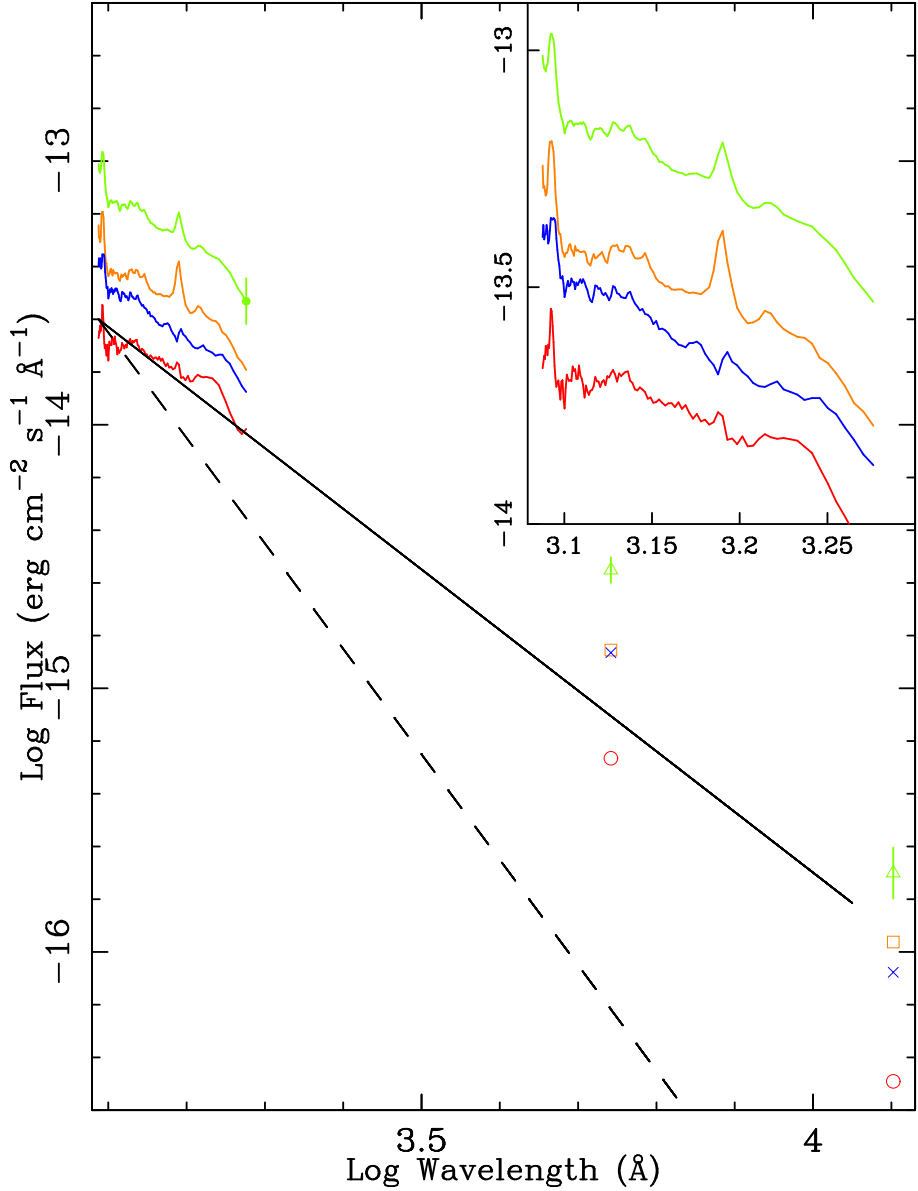}
\caption{The dereddened UV and $V$- and $J$-band fluxes.
  The top (green) line and triangles show the mean fluxes at
  orbital phases 0.75 -- 0.85 and 0.15 -- 0.25, during which
  the system is uneclipsed and contribution from the 
  irradiated face of the secondary star is minimized.
  The next (orange) line and squares show the fluxes during
  eclipse.
  The third (blue) line and crosses are the green minus orange 
  fluxes, giving
  the flux from that part of the system that is eclipsed.
  The bottom (red) line and open circles show the mean fluxes
  at phases 0.25 -- 0.75 minus the uneclipsed (green) fluxes, which should
  be dominated by the flux from the irradiated face of the 
  secondary. 
  The solid line shows $f_\lambda \propto \lambda^{-7/3}$ and
  the dashed line $f_\lambda \propto \lambda^{-4}$.
}
\label{SEDfig}
\end{figure}

Figure~\ref{SEDfig} shows de-reddened UV and $V$- and $J$-band 
fluxes calculated for $E(B-V) = 0.15$ and the extinction law
of \citet{car89} for $R_V=3.1$.
The top (green) line and triangles in the figure show the mean 
fluxes over orbital phases 0.75 -- 0.85 and 0.15 -- 0.25, 
during which the system is uneclipsed and contribution from 
the irradiated face of secondary star is minimized.
The next (orange) line and squares show the fluxes during
eclipse.
The third (blue) line and crosses are the green minus orange 
fluxes, giving
the flux from that part of the system that is eclipsed.
All three of the SEDs
roughly follow an $F_\lambda \propto \lambda^{-7/3}$
distribution from the UV to the $V$-band, and then fall
significantly below $\lambda^{-7/3}$ in the $J$-band.
The photometry of \citet{mas82b} shows that
the SED steepens further to a blackbody $\lambda^{-4}$
slope in the $H$ and $K$ bands.
While it is tempting to identify the $\lambda^{-7/3}$ 
SEDs with the 
theoretical SED for an optically-thick, steady-state $\alpha$-model 
accretion disk, this identification is probably not correct.
It is difficult to make an $\alpha$-model disk produce the
same SED at mid-eclipse, when all but the outer parts of the
disk are obscured, as it produces when it is not eclipsed
and the entire disk is visible. 
Moreover, as we will show in the next section, most of
the disk is hidden by optically-thick vertically-extended
structures and is invisible at UV/optical wavelengths,
so the theoretical SED of a simple $\alpha$-model disk
is not relevant.

The bottom (red) line and open circles in Figure \ref{SEDfig} show 
the mean fluxes at phases 0.25 -- 0.75 minus the uneclipsed (green) fluxes.
This red SED isolates the flux in the large hump in
the light curve at orbital phases opposite the eclipse.
The flux contributed by the irradiated face 
of the secondary star is maximized at these phases.
The flux from the irradiated secondary should be
roughly a multi-temperature blackbody distribution.
The red SED appears to be flattening near
$\log \lambda = 3.1$, suggesting that the flux-weighted
temperature of the heated face is $\sim 25,000$~K.

With the exception of the \ion{N}{5} line, the flux in 
the UV emission lines is neither reduced during eclipse nor
enhanced at other orbital phases.
This shows clearly in the spectrum of the eclipsed parts 
of the system (the blue spectrum), which is nearly devoid of 
emission lines.
The emission lines must, then, arise from optically-thin material
far above and below the disk.
After accounting for the extra width introduced by the poor 
spectral resolution of the UV spectrogram, we find the full
width at zero intensity of the \ion{C}{4} emission line
to be $\sim 45$~\AA.
We compare this to its width in the {\it HST} Faint Object Spectrograph spectrograms from which the data presented by 
\citet{puc95} were derived.  In these spectrograms the width is near 40~\AA.
These widths correspond to projected velocities up to 
$\pm 4400$~km~s$^{-1}$.
Thus the lines come from a high-velocity wind.
Neither the strength nor the
width of the \ion{C}{4} line depend on orbital phase. 

Since the material in the wind must be continuously 
replenished by outflowing gas closer to 
the disk, there must be a base to the wind.
We do not observe \ion{C}{4} emission from the base 
of the wind, so either the base does not produce 
\ion{C}{4} emission or it is hidden by 
vertically-extended optically-thick material further out 
in the disk.
The N~{\small V} line is an exception.
Its strength decreases by $\sim 50$\% during eclipse and
arises nearer the orbital plane.

\section{Light Curve Analysis and a Model for 4U~1822--371}

\subsection{Preliminaries}

\subsubsection{The Light Curve Synthesis Program}
We modeled the UV/optical light curves of 4U~1822--371 using
a light curve synthesis program\footnote{A full description of the program is available at \texttt{http://pisces.as.utexas.edu/robinson/XRbinary.pdf}.}
 we wrote to model X-ray binaries
(e.g. UW Cor Bor, \citealt{mas08}) 
and related objects (e.g. SS Cygni, \citealt{bit07}).  
The program assumes that the orbits of the stars are circular,
the secondary star fills 
its Roche Lobe, and the primary star is a point source 
surrounded by an accretion disk.
The accretion disk can have a complicated geometry.
It can be non-circular,
non-axisymmetric, and vertically extended; and it can have a
rim, an interior torus, and bright or dark spots.
The flux from the secondary star has a realistic spectrum using Kurucz stellar atmospheres \citep{kur96}
but the remaining parts of the system emit like a blackbody.
The program includes heating by irradiation.
Because the geometry and temperature distributions in 
the model can be complex and asymmetric, the program 
resorts to ray tracing to calculate both irradiation 
and the orbital light curve.
The program outputs synthetic light curves for 
Johnson/Cousins filter bandpasses and for square bandpasses 
over user-specified wavelength ranges. 
If provided with an observed light curve, our code calculates
the $\chi^2$ for the fit of the synthetic light curve
to the observed light curve.  A calling program allows the user to find the model parameters
that minimize $\chi^2$ either by a grid search
through parameter space or by a simplex algorithm.

\subsubsection{The Masses and Dimensions of 4U1822--371}
The arrival times of the X-ray pulses from the neutron star
yield accurate values for the projected 
semi-major axis of the neutron star orbit 
$a_{NS}\sin i = (3.015\pm0.015) \times 10^{10}\ \textrm{cm}$
and the mass function 
$f(m_2)= m_2 q^2 (1 + q)^{-2}\sin^3 i=0.0203\pm0.0003$
\Msun, where $a_{NS}$ is the radius of the neutron star's
orbit and $i$ is the orbital inclination \citep{jon01}.
The orbital eccentricity is not measurably different from 
zero.

It appears to us that the masses of the two stars
in 4U~1822--371 have not yet been determined with certainty.
The mass function can be recast to the form
$m_{NS} \sin^3 i = (0.0203\pm0.0003)q^{-3}(1+q)^2\ M_\odot$.
The eclipse solution limits the orbital inclination to a 
narrow range near $82^\circ$, so
$m_{NS} \approx 0.0209q^{-3}(1+q)^2\ M_\odot$.
There have been several attempts to determine the
mass ratio from the radial velocity curve
of the secondary star.
The spectrum of the heated face of the secondary 
star has \ion{He}{1} absorption lines and a narrow
emission line from \ion{N}{3}~$\lambda 4640$, which 
is excited by the Bowen fluorescence mechanism 
\citep{har97, cas03}.
The radial velocity curve of the \ion{He}{1} 
absorption has been measured and its
amplitude is $K_{HeI} \approx 230\ \textrm{km s}^{-1}$
\citep{har97, cow03, cas03}.
It is, however, unclear how to interpret $K_{HeI}$ because 
the absorption lines arise on only one face of the 
secondary and because the line profiles are probably 
distorted by helium emission lines from the accretion disk.
The radial velocity curve also suffers from a phase 
shift of $\sim 0.1P_{orb}$ with respect to the 
pulsar ephemeris, further complicating its
interpretation \citep{jon03, cas03}.

The radial velocity curve of the \ion{N}{3}~$\lambda 4640$
emission line also has been measured and has an 
amplitude $K_{NIII} = 280 \pm 3\ \textrm{km s}^{-1}$
\citep{cas03,mun08}.
We agree with \citet{cas03} that this is
a firm lower limit to the amplitude of the secondary's radial 
velocity curve, which translates to lower limits on the 
masses:
$m_{NS} \ge 1.14\ M_\odot$ and $m_2 \ge 0.36\ M_\odot$.
Correcting from $K_{NIII}$ to the true amplitude $K_2$ 
of secondary star's radial curve requires a model
for the distribution of the \ion{N}{3}~$\lambda 4640$
emission across the face of the secondary star.
\citet{mun05,mun08} have developed such a model and 
used it to deduce $1.52\ M_\odot < m_{NS} < 1.85\ M_\odot$.
In their model the \ion{N}{3}~$\lambda 4640$ line is
emitted uniformly over the face of the irradiated secondary
except where shadowed by a simple thin disk with a raised, 
optically-thick rim.
We will show that a more complex model is required for  
the disk in 4U~1822--371.  
The K-correction derived from the thin disk model is likely be
only qualitatively correct.

As the masses of most neutron stars in binary X-ray pulsars fall in a narrow range
near $1.35 \Msun$ \citep{tho99, nic08}, we adopted this value for 
most of our models.
The mass function then becomes a relation between $q$ and $i$,
materially reducing the parameter space that needs to
be explored when fitting models.
This choice of neutron star mass also restricts other properties of 
4U1822--371 to lie in a narrow range:
If $i$ lies in the range $78^\circ < i < 84^\circ$, then
$0.2948 < q < 0.3007$, $0.398 < m_2/M_\odot < 0.406$,
and the separation of the centers of mass of the two stars
is $1.3336 < a/10^{11}\textrm{cm} < 1.3367$.
Since the secondary star must fill its Roche lobe, its
radius is close to $0.54\ R_\odot$, placing it close to
the observed main-sequence mass-radius relation
\citep{dril00}.
The temperature of a main-sequence star with this mass and radius
is $\sim 3500$~K.
This is a low enough temperature that we ignore flux from
the secondary star except from those parts of its surface
that are heated by irradiation.

Because \citet{mun08} argue for a high-mass neutron star
in 4U~1822--371, we have also explored the effect of neutron
star mass on our results, adopting the extreme value
$m_{NS} = 2.0\ M_\odot$ for a few models.
For $m_{NS} = 2.0\ \Msun$ and $i = 81^\circ$ the mass function
yields $q = 0.255$, $M_2 = 0.51\ \Msun$
and $a = 1.50 \times 10^{11}\textrm{cm}$.
The radius of the secondary star is $0.58\ R_\odot$, again
placing it close to the main-sequence mass-radius relation.
A main-sequence star with a mass of $0.51\ \Msun$ has
a temperature near $\sim 3800$~K.
Fits of these higher-mass models to the light curves yielded
small quantitative differences in the fitted parameters 
but the overall results were qualitatively almost identical.
The derived orbital inclinations, for example, were typically 
higher by $\sim 0.75^\circ$ and the heights of the disk 
rim increased by 10-15\%.
This also means, though, that the mass of the neutron star
is not constrained by the optical/UV light curves.
We report results only for the $m_{NS} = 1.35\ M_\odot$
models.

\subsubsection{The Accretion Disk}
As discussed in the introduction, the unobscured X-ray
luminosity of 4U~1822--371 is high, possibly near
Eddington; and the high mass flow rates deduced from
the rapidly changing orbital period also suggest a luminosity
near Eddington.
Since the X-ray luminosity of 4U~1822--371 is relatively 
constant in the sense that no large-amplitude outbursts 
have been observed, the accretion disk in 4U~1822--371 is 
at least roughly in a steady state.
These considerations suggest strongly that the accretion disk is
optically thick and its inner edge is near the neutron star.

While we will not necessarily limit our disk models 
to $\alpha$-model disks,
an optically-thick $\alpha$-model disk is the default model 
for the outer parts of the disk in 4U~1822--371.
The maximum radius of the disk is thought to
be set by viscous dissipation induced by tidal 
interactions with  the secondary star.
For a mass ratio $q = 0.3$ the tidal
truncation radius is 
$r_{\rm disk} \approx 0.43 a  \approx 5.7 \times 10^{10}$~cm \citep{fkr}.
Let us calculate a typical temperature and thickness of the disk at a radius that is large, but less than the radius of its outer edge.
For $\alpha = 0.1$, $m_{NS} = 1.35\ M_\odot$,
a mass flow rate of
$10^{-8}\ M_\odot\ \textrm{yr}^{-1}$, and a
radius of $r = 4 \times 10^{10}$~cm 
an optically-thick steady-state $\alpha$-model disk has
$h/r_{\rm disk} =  0.04$ and $T_{eff} = 7800$K \citep{fkr}.
These values depend only weakly on $\alpha$, $m_{ns}$,
and $\dot m$.
Except where it is hit by the 
stream of transferred material from the secondary star, 
the outer, vertical rim of the disk should not be even 
as hot as this low temperature. 
We will see that the visible parts of the 4U~1822--371
system have a brightness temperature near $3 \times 10^4$~K
at UV/optical wavelengths.
The temperature of the rim is so much lower that it
is a minor contributor to the flux.
If the rim is vertically extended, its inner surface may be
heated to higher temperatures but this will only
emphasize the low surface brightness of the outer surface of
the rim.

As a result, our models emit negligible flux from the disk rim.
In the previously published models for the UV/optical light curve
of 4U~1822--371, the disk rim is a significant contributor, 
often the dominant contributor to the UV/optical flux.
This is shown most clearly in Figure~9 of \citet{hel89}.
Many of the differences between our results and earlier
results flow from the different handling of the disk rim
flux.

\subsection{The Disk Is Large, Its Center Obscured}
\begin{figure}
\center \includegraphics[scale=0.65, angle=270]{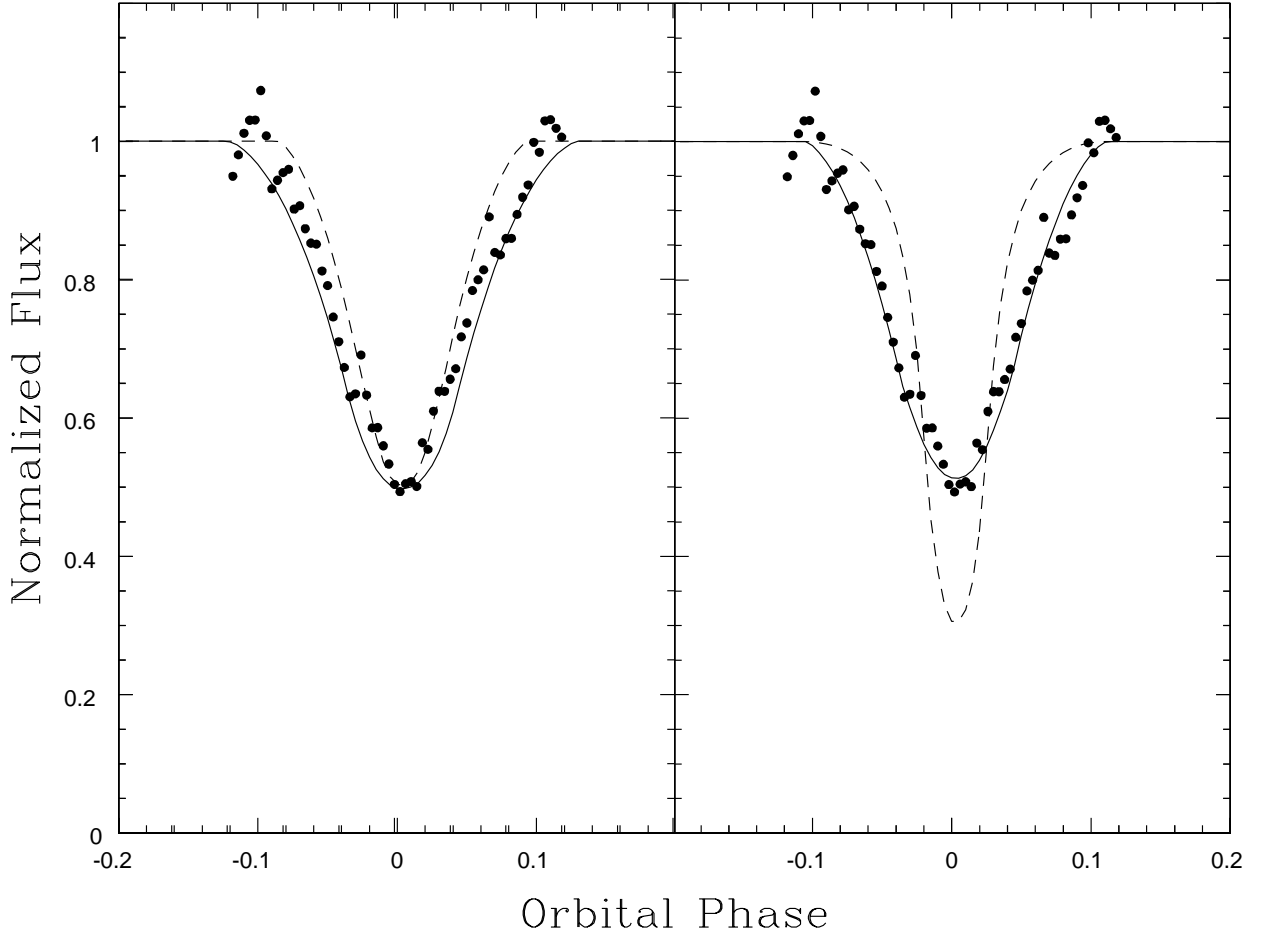}
\caption {The solid circles show the co-added and binned UV eclipse light
curve of 4U~1822--371 normalized to 1.0 just outside
the eclipse. {\it Left Panel:} The lines are synthetic 
light curves for a geometrically-flat disk with a
flat temperature distribution ($T\propto r^{-0.1}$)
and for an orbital inclination of $81^\circ$.
The solid and dashed lines are the light curves for 
disks with $r_{\rm disk}/a = 0.45$ and 0.30 respectively;
both have been normalized to emphasize the eclipse shape.
The disk in 4U~1822--371 must be large, $r_{\rm disk}/a \approx 0.40$, with
its precise radius depending on the orbital
inclination and other disk properties. {\it Right Panel:} The dashed
line is the eclipse light curve for a  geometrically-flat disk with an
$\alpha$-model temperature distribution ($T\propto r^{-3/4}$) and a
surface temperature of 20,000~K at the maximum radius, but 2,000~K at the
outer edge.  The solid line is the synthetic light curve for a disk
with the same surface temperature at its maximum radius but a nearly flat
temperature distribution ($T\propto r^{-0.1}$).  The synthetic light
curves have been normalized to emphasize the shape of the eclipses.
The $\alpha$-model disk produces an eclipse that is much too narrow to
fit the observed eclipse.
\label{FigureAB}}
\end{figure}

We first establish that the accretion disk is 
large and its center obscured at UV/optical wavelengths. The left panel in
Figure~\ref{FigureAB} shows the eclipse portion of the
binned UV light curve normalized to 1.0 just outside eclipse.
The eclipse is wide, lasting $\Delta \phi \approx 0.20$ in phase.
The panel also shows two synthetic light curves, both 
calculated for a disk that is
geometrically flat, has a nearly flat temperature distribution
($T \propto r^{-0.1}$), and is eclipsed at an orbital
inclination $i = 81^\circ$.
One light curve was calculated for $r_{\rm disk}/a = 0.30$, 
the other for $r_{\rm disk}/a = 0.45$, and both
have been normalized to match the flux at mid-eclipse to 
emphasize the effect of disk width on the light curve.
Disks with $r_{\rm disk}/a \gtrsim 0.40$ are needed 
to fit the observed eclipse.

The dashed line in the right panel of Figure~\ref{FigureAB} is the synthetic
light curve for a geometrically-thin, optically-thick, steady-state 
$\alpha$-model disk with a $T\propto r^{-3/4}$ 
temperature distribution.
The radius of the disk is $r_{\rm disk}/a = 0.4$, its temperature
is 20,000~K at the outer edge, and the orbital inclination
is $77^\circ$.
The bright central regions of this model disk produce
an eclipse light curve with a deep, narrow  
center in striking disagreement with the observed eclipse.
The solid line in the panel is the synthetic
light curve for a disk model with similar geometry but with 
a flatter temperature distribution to reduce the 
brightness of its central regions
($T \propto r^{-0.1}$ and $i = 78.5^\circ$).  
This model matches the general shape of the observed 
eclipse and, indeed, the fit is formally excellent.
We do \emph{not} conclude that the disk actually has a
flat temperature distribution as we consider a flat
distribution physically unrealistic.
We do conclude that the flat temperature distribution
is mimicking the true appearance of the disk, a disk that
appears not to have bright central regions.
As argued in the previous section, the central regions of the
disk must be hot and bright, so the central regions 
of the disk must be obscured by disk structures 
extending above and below the orbital plane.

We now consider two broad groups of models for obscuring
the central disk.
The first group includes models 
similar to standard thin-disk models.
In these models the edge of the disk is tall enough 
to hide the center of the disk either because the disk 
has a large flare or because the disk is flat but 
has a vertically extended rim.
In the second group of models the obscuration is produced
by vertically-extended structures interior to the disk edge.

\subsection{Obscuration by a Tall Disk Rim?}
\begin{figure}
\center \includegraphics[scale=0.65]{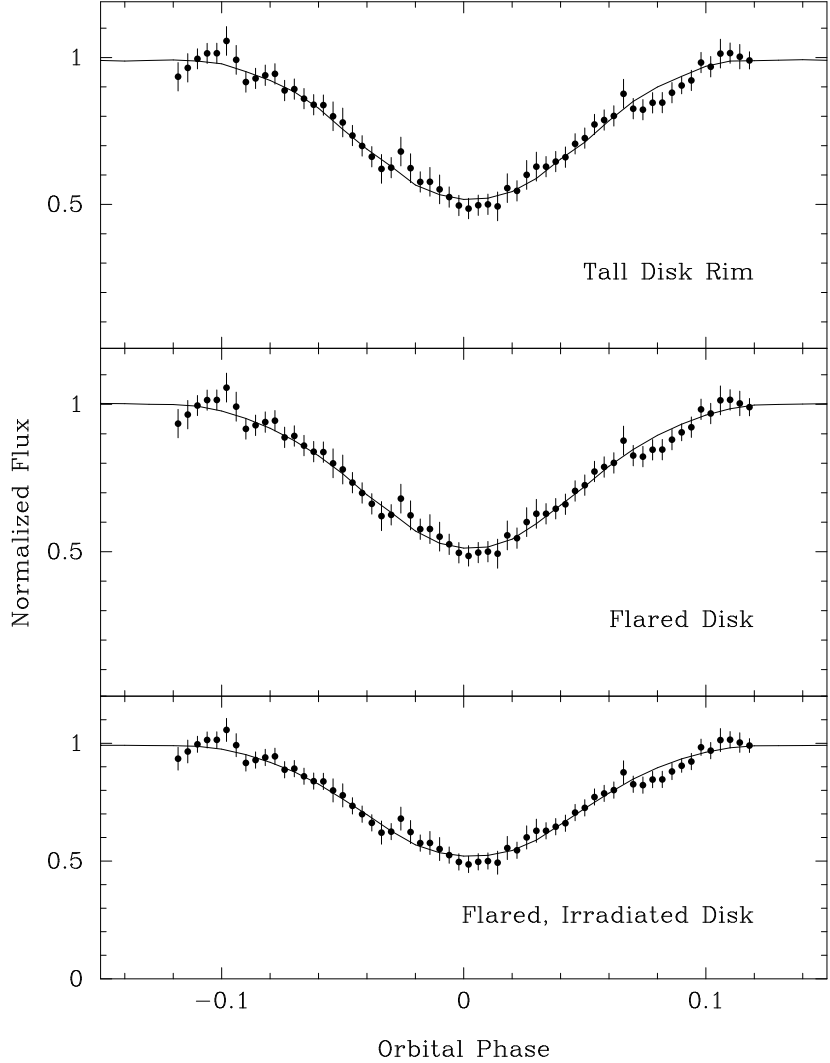}
\caption{ The solid circles in all three panels
show the co-added and binned UV eclipse light 
curve of 4U~1822--371 normalized to 1.0 just outside the eclipse.
The solid line in each panel is a synthetic light curve
fitted to the observed eclipse.
Although the synthetic light curves fit the data well,
they are all fatally flawed.
The central regions of the disk are unobscured,
so the models must have a flat temperature distribution, 
thus failing to avoid the very problem they were contrived 
to solve. 
The neutron star is also visible, 
disagreeing with X-ray observations.
{\bf Top Panel:}
The synthetic light curve for a thin-disk tall-rim model.  
The orbital inclination of the model is $81^{\circ}$,
the disk radius is $r_{\rm disk}/a=0.44$, and 
the rim height height is $h_{\rm rim}/a=0.056$.
{\bf Middle Panel:}
The synthetic light curve for a concave
flared disk.
The orbital inclination of the model is $78^{\circ}$,
$a_{max}/a=0.43$, $H_{\rm edge}/a=0.060$, and $H_{pow}=2.0$ 
(see equation~\ref{FlareDisk-eq}).
{\bf Bottom Panel:}
The synthetic light curve for a concave, flared, 
irradiated disk.
The orbital inclination of the model is $81^{\circ}$,
$a_{max}/a=0.45$, $H_{\rm edge}/a=0.060$, and $H_{pow}=2.0$.
The disk is irradiated by flux from the neutron star.
}
\label{TallRimModels}
\end{figure}

The top panel of Figure~\ref{TallRimModels} shows the observed 
UV eclipse light curve 
and a synthetic light curve produced by a
geometrically-thin disk with a vertically-extended rim.
The orbital inclination of the model is $81^{\circ}$,
the disk radius is $r_{\rm disk}/a=0.44$, and 
the rim height is $h_{\rm rim}/a=0.056$, or $h_{\rm rim}/r_{\rm disk} \approx 0.13$.
This rim is quite tall, raising concerns about the physical 
validity of the model; but the tall rim does produce a 
light curve that fits the observed light curve fairly well.
Nevertheless, this model is fatally flawed.
It has a combination of rim height and orbital inclination 
that leaves most of the central disk unobscured.
Therefore, to avoid an eclipse that is too narrow and deep, 
the model must have a flat temperature distribution, 
thus failing to avoid the very problem it was contrived 
to solve. 
Worse, the neutron star is visible, 
disagreeing with X-ray observations.
This is a generic and robust result for thin-disk
tall-rim models.
We have explored the parameter space for these models 
extensively, 
varying the orbital inclination, the disk radius and
temperature, the rim height and temperature,
and the irradiation of the
inner surface of the rim by the inner disk and neutron star.
Models that fit the observed light curve with an acceptable 
$\chi^2$ always have a flat temperature distribution 
and the neutron star is usually visible.

Flared disks -- disks whose thickness increase with radius -- 
automatically have thick rims that can hide the inner disk 
and neutron star.  In our computer code the 
height of a flared disk above the orbital plane
is given by
\begin{equation}
H = H_{edge}\left(\frac{a-a_{\rm min}}{a_{max}-a_{\rm min}}\right)^{H_{pow}},
   \label{FlareDisk-eq}
\end{equation}
where $a_{\rm min}$ and $a_{max}$ are the radii of the inner and 
outer edges of the disk, and $H_{\rm edge}$ is the thickness of the
disk at its outer edge.
If $H_{pow} >1$, the disk has a concave flare.
The middle panel of Figure~\ref{TallRimModels} shows a 
synthetic light curve for a flared disk.
The model disk has $H_{\rm edge}/a=0.060$, $a_{\rm min}/a=0.01$, 
$a_{max}/a=0.43$, and $H_{pow}=2.0$.
The synthetic light curve fits the observed light curve 
about as well as the light curves for thin
disks with thick rims but 
it has the same fatal flaws: The disk has a flat temperature 
distribution and the neutron star is usually visible.

Models with concave flared disks have an important advantage
over those with flat disks:  The outer regions of the disk can 
be heated by irradiation from the neutron star and inner disk.
The heating flattens the temperature distribution, providing
a natural explanation for the flat distribution.
The bottom panel of Figure~\ref{TallRimModels} shows
a synthetic eclipse light curve for a flared disk irradiated by the neutron star.
The orbital inclination is $81^{\circ}$,
$a_{max}/a=0.45$, $H_{\rm edge}/a=0.060$, and $H_{pow}=2.0$.  This value of $H_{pow}$ is 
much larger than the standard $H_{pow}=9/8$ for an $\alpha$-model disk,  
but lower values of $H_{pow}$ did not fit the observed light curve well.
The fit for this particular model is good but the neutron star remains visible.

In a final attempt to use the disk rim to block light from 
the center of the disk, we invoked a concave flared disk
with an additional vertically-extended rim and allowed
the disk and inner face of the rim to be irradiated by 
the neutron star.
This model has enough free parameters to produce an
excellent fit to the observed light curve at all orbital
phases, not just the eclipse.
But the inner disk and the neutron star still
remain visible and the model fails.
Thus, we reject this entire group of models.

\subsection{Torus Plus Variable-Height Rim}
The previous discussion leads us to consider vertically-extended
disk structures closer to the neutron star.
Inspired by the presence of the ADC at X-ray wavelengths,
we give the disk vertical extension by adding an opaque
torus (a donut with the neutron star centered in
the hole of the donut).
Our code allows the torus to have an elliptical cross section 
and specifies its geometry by three parameters:
the distance from the center of the cross-section of the torus to the
neutron star, which we call its radius or $a_o$,
the maximum height of the torus above and below the orbital plane, 
and the width of the torus between its inner and outer walls
as measured in the orbital plane.
The toroidal structure should not be taken too literally.
Because 4U~1822--371 has a high orbital inclination, we observe the
torus from the side and, like a donut viewed from the side,
we observe only the outer wall of the torus.
The observations give little information about the inner
wall or about the interior of the torus.
Furthermore, despite our original motivation for including
the torus, we will see that the properties we derive for 
the torus are not consistent with the properties of an ADC
and we will instead identify the torus with the 
optically-thick base of the disk wind.

We assume that the torus emits blackbody radiation and
for simplicity we give the surface of the torus
a single temperature except where it is heated by irradiation. The temperature of the torus is poorly constrained.
The SED of 4U~1822--371 is not a blackbody distribution,
which precludes assignment of a meaningful color
temperature. 
The large uncertainty on the distance and reddening 
to 4U~1822--371 preclude measurement
of an accurate brightness temperature, but the
brightness temperature of the outer wall of the torus
must be surprisingly low.
Fits to the UV eclipse show that the outer wall
extends to $\sim 0.35a$ in the orbital plane 
($\sim 80$\% of the disk radius)
and the maximum 
height of the torus is $\sim 0.2a$ ($\sim 50$\% of the
disk radius), giving the torus 
a projected surface area of roughly
$0.28a^2 \approx 4 \times 10^{21}\ \textrm{cm}^2$.
About half the projected area is eclipsed, 
$A_{ecl} \approx 2 \times 10^{21}\ \textrm{cm}^2$,
and the de-extincted eclipsed flux is 
$f_{ecl}=3.1 \times 10^{-14}$~erg/s/cm$^{2}$/\AA\
at 1600 \AA\ (the blue line in Figure~\ref{SEDfig}).
The brightness temperature of the eclipsed surface of the 
torus can be calculated from
$B_{\lambda}= f_{ecl} D^2/A_{ecl}$,
where $D$ is the distance to 4U~1822--371.
For $D = 5000$~pc
the monochromatic brightness temperature at 1600~\AA\
is $T_b \approx 26,000$~K.   
Although this temperature cannot be trusted to within a 
factor of two or so, it is $\sim10^3$ times lower than 
temperatures typically associated with coronae.
Thus, the brightness temperature of the torus -- at least the outer wall
of the torus -- is not consistent with an ADC.
The temperature is, however, high enough that the torus
dominates the UV/optical flux from 4U~1822--371.
This justifies our earlier claim that the flux
from unirradiated surfaces of the secondary star 
and the outer surface of the disk rim can be ignored.

This torus-like structure is far too tall 
to agree with the thickness of an $\alpha$-model disk 
for any reasonable choice of parameters.
Its temperature is far too low for it to be part of the
ADC.
What, then, is this vertical structure?
Since we have already found a high velocity disk wind
and have shown that there is much mass loss from the
disk, we suggest that the vertical structure is simply
the optically-thick base of the disk wind.

\begin{figure}
\center \includegraphics[angle=270,scale=0.65]{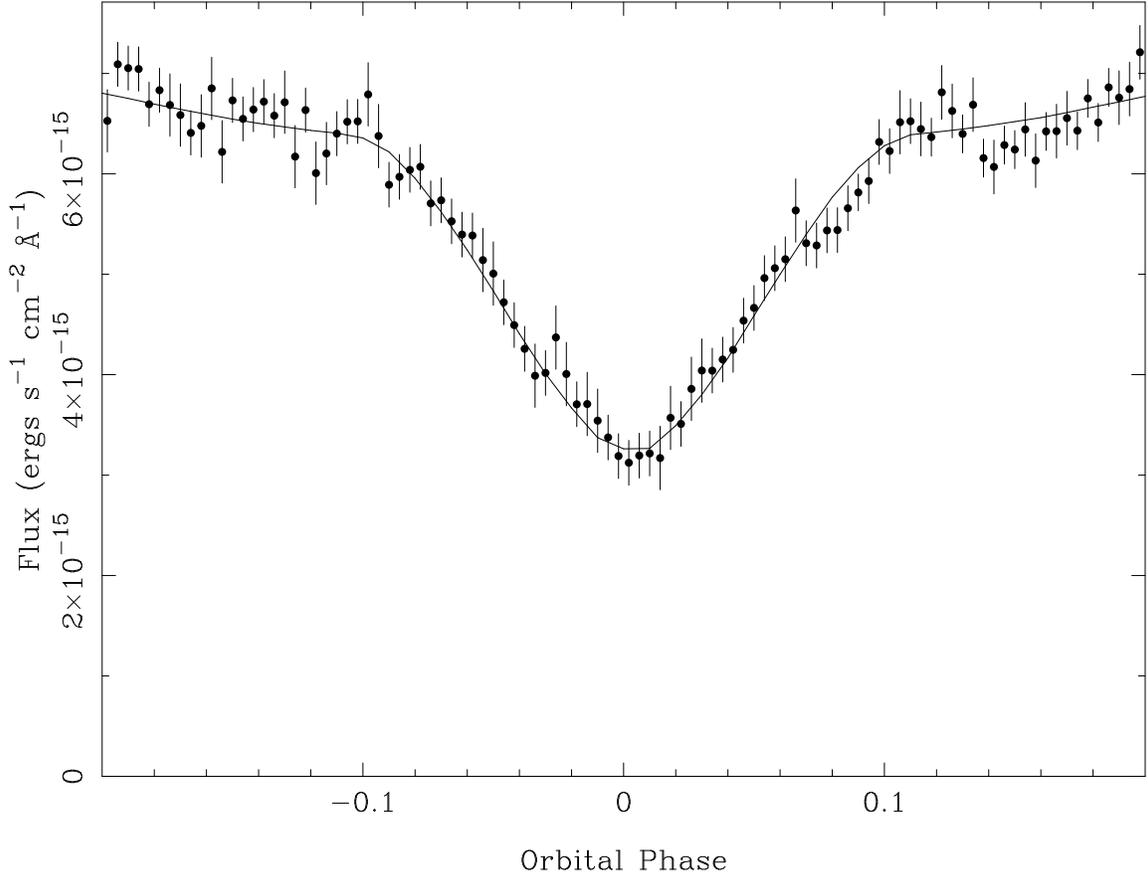}
\caption{The solid circles show the observed UV eclipse light 
curve of 4U~1822--371.
The solid line is a synthetic light
curve fitted to the data.
The model from which the light curve was calculated
has an orbital inclination of $83.5^\circ$, 
a concave flared disk with radius $r_{\rm disk}/a=0.43$
and an edge height $H_{\rm edge}/r_{\rm disk} = 0.07$.
The disk is covered by an opaque torus with midpoint at $a_o=0.53r_{\rm disk}$,
maximum height of $0.56r_{\rm disk}$, and width of $0.56r_{\rm disk}$, so that the
inner wall of the torus extends in to $0.26r_{\rm disk}$ and the
outer wall out to $0.81r_{\rm disk}$.
The torus blocks flux from the neutron star and from
the disk out to $0.81r_{\rm disk}$,
and the edge of the flared disk blocks flux from most 
of the rest of the disk.
}
\label{TorusEclp}
\end{figure}

The torus by itself does not produce a synthetic 
light curve that fits the observed eclipse light curve 
adequately,
but the addition of either an opaque vertically-extended rim 
or a flared disk yields an excellent fit.
The best-fit synthetic light curve for a flared 
disk is shown in Figure~\ref{TorusEclp}.
The model from which the light curve was calculated
has an orbital inclination of $83.5^\circ$, 
a concave flared disk with radius $r_{\rm disk}/a=0.43$
and edge height $H_{\rm edge}/r_{\rm disk}=0.03$.
The midpoint of the torus is at $a_o = 0.53r_{\rm disk}$, its
maximum height is $0.56r_{\rm disk}$, and its width is $0.56r_{\rm disk}$, so that the
torus extends from $0.26r_{\rm disk}$ to $0.81r_{\rm disk}$.
The torus blocks flux from the neutron star and all the disk out
to $0.81r_{\rm disk}$,
and the edge of the flared disk blocks flux from most 
of the rest of the disk.
Thus the torus solves the two problems that vitiated
models with flat or flared disks and raised rims.

\subsection{A Model for the Entire Orbital Light Curve}

We now model the entire orbital light curve using the
eclipse model from Section 5.4 as a starting point.
The salient feature of the non-eclipse part of the orbital
light curve is the large-amplitude, roughly-sinusoidal hump.
The phase of the hump might lead one to suspect that 
the hump is caused primarily by the varying visibility of the 
irradiated face of the secondary star but our attempts 
to model the hump with just 
the irradiated secondary star yielded poor fits.
The hump's precise phase, its shape, and the variations of
its shape from year to year cannot be produced by the
irradiated secondary alone.

The model producing the synthetic light curve shown in 
Figure~\ref{TorusEclp} had a torus and a disk with 
a dark, vertically-extended edge.
We now allow the height of the edge to vary with 
azimuth around the neutron star.
As its height varies, the rim blocks more or less of
the flux from the torus, helping to produce the hump.
In the previous section we introduced the vertically-extended
edge by invoking a concave flared disk, but
a flat disk with a vertically extended rim would have
worked just as well.
In the current context it is more convenient to
model the edge with a flat disk and extended rim,
allowing the rim height to vary with azimuth. 

We also include flux from the heated secondary in the
model and allow fits to the light curve to determine
how much flux the secondary contributes.
The secondary is heated in part by the torus and disk
but mostly by X-ray flux.
While we include X-ray flux in the model, it is not 
a physically meaningful quantity since the
amount of flux needed depends sensitively on the geometry 
of the X-ray emitting regions, on the radiative transfer of
the flux through the ADC and torus, and on the reprocessing and 
redistribution of the flux in
the secondary star.
As we have little information about any of these factors, we
regard the X-ray flux as merely a convenient parameter 
to specify the irradiative heating of the secondary star.
With the addition of a variable-height rim and flux from
the irradiatively-heated secondary star we
achieved successful fits to the entire
orbital light curve and no further complications were needed.

While the foregoing discussion has been necessarily lengthy,
the resulting model is actually rather simple.
The parameters describing the geometry of the model are
\vspace{-0.5\baselineskip}
\begin{itemize}
\vspace{-0.5\baselineskip}
\item The orbital inclination, which
      also fixes the mass ratio and the mass of the secondary
      star through the mass function,
\vspace{-0.5\baselineskip}
\item the outer radius of the disk,
\vspace{-0.5\baselineskip}
\item the height, width, and radius of the disk torus, and
\vspace{-0.5\baselineskip}
\item the height of the disk rim as a function of azimuth
      around the neutron star.
\vspace{-0.5\baselineskip}
\end{itemize}
\vspace{-0.5\baselineskip}
We assume that the secondary star and the disk rim have low
temperatures and do not contribute significant flux except
where heated by irradiation.
Fits to the light curve show that
the inner regions of the disk are hidden by the torus
and the outer regions are hidden by the disk rim, so the
disk is not visible and its temperature is irrelevant.
Thus the only temperature that needs to be specified is the
monochromatic brightness temperature of the torus.
Finally, the amount of X-ray energy irradiating the 
disk rim and the secondary star must be specified.
There are also a few nuisance parameters such as the
zero point in orbital phase, a scale factor on the
flux, and various albedos that must be specified
but are irrelevant to the model.  

\begin{figure}
\center \includegraphics[angle=270,scale=.65]{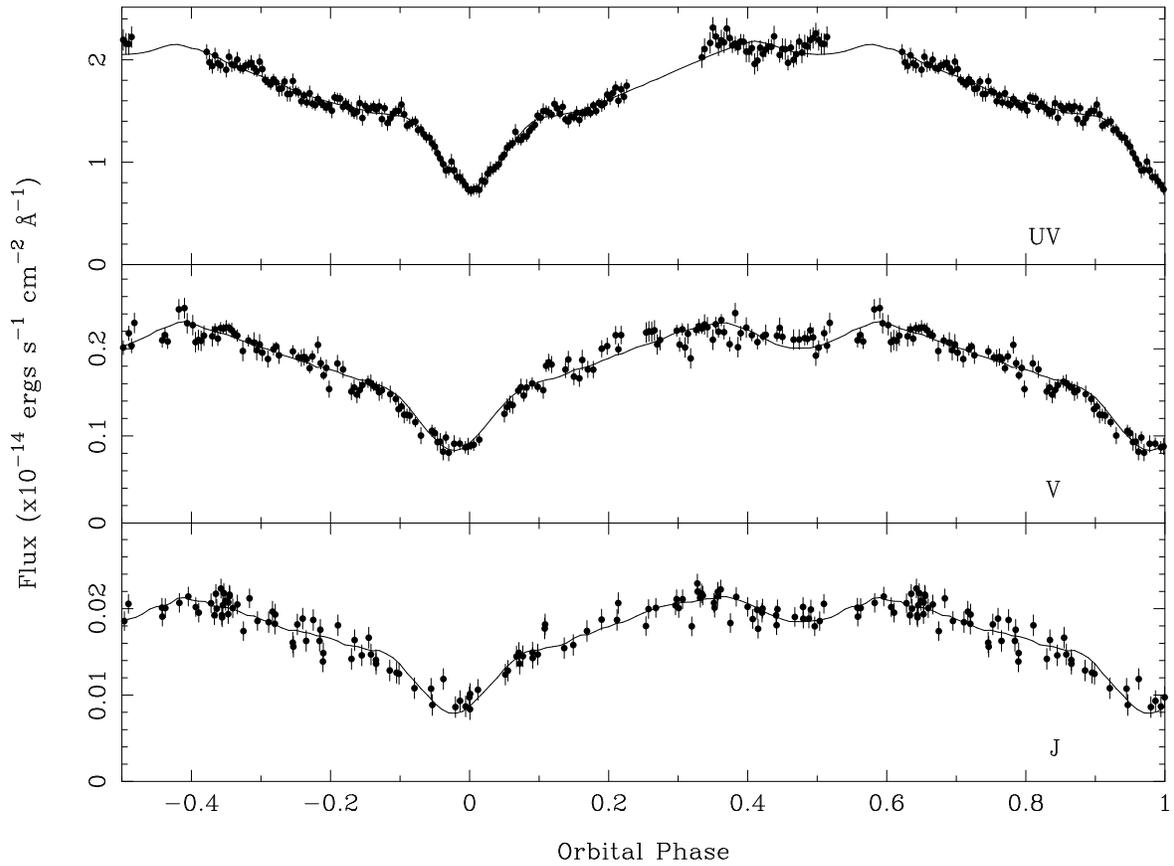}
\caption{The best fitting UV, $V$, and $J$ light curves with the best
  fitting model in Table \ref{model}.  This model includes an optically thick torus.}
\label{3lc}
\end{figure}

\begin{figure}
\center \includegraphics[angle=270,scale=.65]{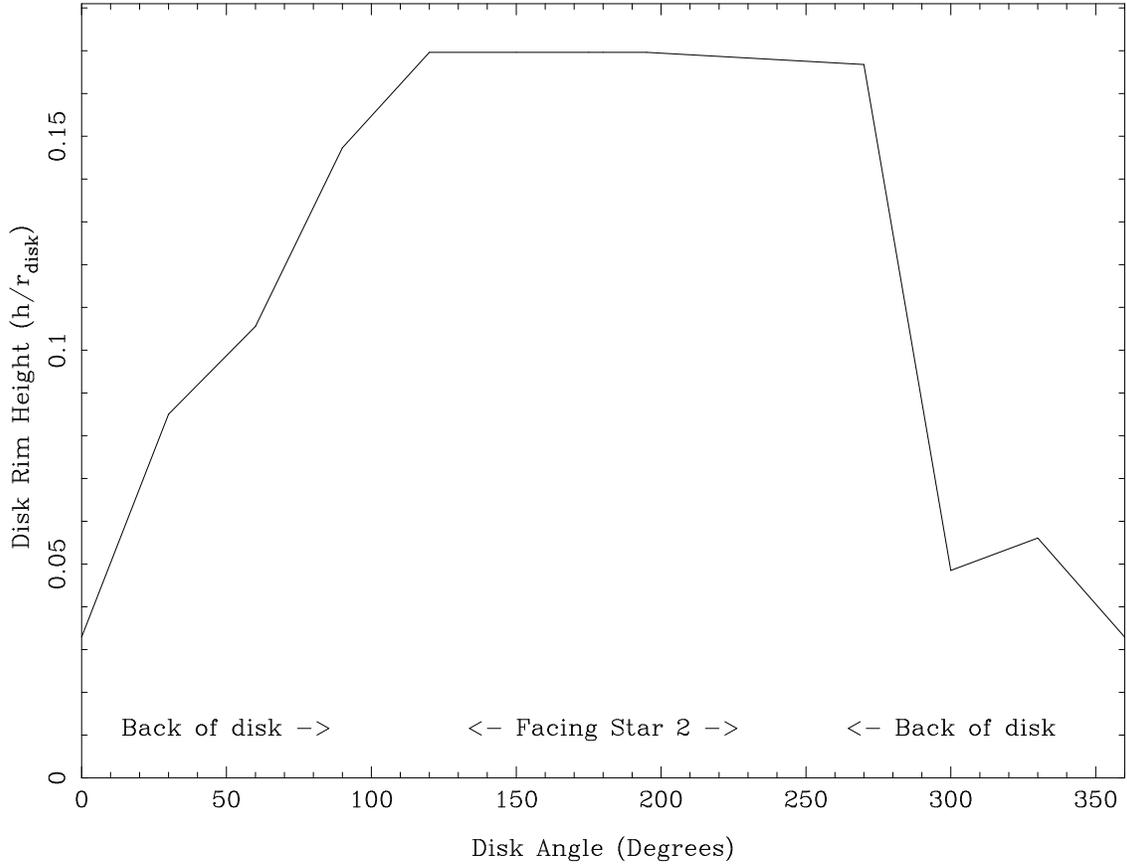}
\caption{The variation in disk rim height with azimuthal disk angle. An angle
  of $0^{\circ}$ is away from the star, rotating opposite the direction of
  orbital motion, making $180^{\circ}$ the radius in the disk connecting the
  NS and secondary star.  The rim height is in terms of disk radius and is the height added to the very thin ($0.002r_{\rm disk}$) disk. 
\label{dHt}}
\end{figure}

We fit the model to the observed UV light curve using a
simplex algorithm to find the set of parameters that
minimized the $\chi^2$ of the fit.
We also examined parameter space extensively with grid
searches to test the uniqueness of the fit.
Within the context of the model the values of the fitted parameters are unique
and robust.
The resulting synthetic light curve is superimposed on
the observed light curve in the top panel of
Figure~\ref{3lc}.
The rim height as a function of azimuth angle is shown in 
Figure~\ref{dHt}, where the azimuth angle is defined to be 
$0^{\circ}$ on the side of the disk opposite the secondary star 
and increases in the 
direction opposite the direction of orbital motion.
We fit only the $V$- and $J$-band light curves obtained
in 2006, the same year as the UV light curves were obtained.
To fit the light curves we retained
the values of all the parameters determined from the fit
to the UV light curve except for those specifying
the geometry and temperature of the torus.
The fitted synthetic light curves in the $V$- and $J$-bands
are shown superimposed on the observed light curves
in the bottom two panels of Figure~\ref{3lc}.
The values of $\chi^2$ for the fits
are 163.5 for 195 data points and thirteen free
parameters in UV, 210.3 for 150 data points and four free 
parameters in $V$, and 130.7 for 116 data points and four 
free parameters in $J$, for a total reduced $\chi^2$ of 1.05.

Values for the fitted parameters are given in
Table~\ref{model}.
The radius of the disk is $r_{\rm disk}/a = 0.40 \pm 0.01$ and
the disk rim can have a height up to 
$h_{\rm rim}/r_{\rm disk} \approx 0.17$, but the height for much 
of the disk is near $h_{\rm rim}/r_{\rm disk} \approx 0.05$, close to the $\alpha$-model $h/r$ values (Section 5.1). 
The height of the torus, $0.20a=0.5r_{\rm disk}$, is the same in all three
passbands, but the position of its outer wall is much
different, $a_o=0.31a=0.78r_{\rm disk}$ in the UV, but $a_o=0.39a=0.98r_{\rm disk}$ in the $V$-
and $J$-bands.
This is a reflection of the obvious difference between
the UV eclipse and the broader, flatter bottomed eclipse at longer wavelengths.

This model also fits the 2005 $V$- and $J$-band light curves, but
with a change to the rim height with disk azimuth.
The maximum disk rim height is approximately the same as the 2006 model with $h_{\rm rim}/r_{\rm disk} \sim 0.15$.  However, the rim is now symmetric about the approximate location of the ``hot spot" where the tidal stream encounters the disk and not the radius connecting the companion star and the neutron star.  A detailed quantitative model is unwarranted 
due to the sparse sampling of the data.  

On considering these results, a new issue arises.
While we and \citet{hel89} before us needed to place the
opaque wall near the edge of the accretion disk to fit
the light curve, neither the
X-ray nor the UV/optical light curves require it to be precisely at
the disk edge.
It could, in fact, be anywhere between the edge of the
disk at $\sim 0.4a$ and the outer wall of the torus in the $V$-
and $J$-bands at $\sim 0.39a$.
Although this is not a large change in position, it allows
a significant re-interpretation of the wall.
We placed the wall at the disk rim -- and we suspect
others placed it at the disk rim -- primarily because there
was no other disk structure to which the wall could be
attached.
With the addition of a disk torus the wall can 
plausibly be made part of the torus, becoming an
opaque belt around the outside of the torus.
The wall now becomes part of the disk wind, not
a disk rim.
This re-interpretation has a considerable
advantage.
When made part of the disk wind, the wall no longer need
be in hydrostatic equilibrium, requiring impossibly 
high gas pressure or supersonic turbulence for support.
It is simply a cooler, darker layer at the base of
the disk wind.

\section{Summary and Discussion}

Combining our new times of eclipses with the previously 
published times, we have 
derived an improved ephemeris for the eclipses 
in the UV/optical light curves of
4U~1822--371 (equation~\ref{Ephemeris}).
The quadratic term in the ephemeris yields a timescale
for a change in orbital period of
${{P} / {\dot P}} = (3.0 \pm 0.3) \times 10^{6}\ \textrm{yr}$.
The deduced rate of mass transfer is large,
probably greater than 
$3 \times 10^{-8}\ M_\odot \ \textrm{yr}^{-1}$.
Mass cannot be accreting onto the neutron star at this
rate without violating the Eddington limit, so much
of the transferred mass must be lost from the system.

We have inferred a strong disk wind from the \ion{C}{4} 
emission line in the UV spectrum.
The wind is axially symmetric and has projected outflow 
velocities up to 4000~km~s$^{-1}$.
The flux in the \ion{C}{4} emission lines does not
decrease during eclipses.
Since regions less than $\sim 0.5r_{\rm disk}$ above the disk are 
at least partially eclipsed, the \ion{C}{4} emission arises
from regions yet further above the disk.
The outflowing material closer to the
disk either does not produce \ion{C}{4} emission
or is hidden by other optically-thick structures.

Much of this paper was devoted to developing a model for the
accretion disk in 4U~1822--371 from fits to the UV/optical orbital 
light curves.
The following properties of the accretion disk appear to be
robustly determined.
The disk is large; the fitted radius is 
$r_{\rm disk}/a = 0.40 \pm 0.01$, close to the tidal truncation radius.  
To avoid an eclipse that is deeper and
narrower than the observed eclipse, the neutron star 
and the central regions of the accretion
disk must be obscured.
The dual requirements that the models must
produce synthetic light curves that fit the observed
eclipse and must obscure the neutron star and central
disk forces most of the obscuring material to be at 
intermediate radii in the disk, not at the edge of the disk.
We modeled the obscuring material as an optically thick
torus. 
The torus extends more than 3/4 of the way to the edge of the
disk at UV wavelengths and nearly all the way to the 
edge in the $V$- and $J$-bands; and
it extends to a height of $\sim 0.5r_{\rm disk}$ above
the disk at all wavelengths.

The surface brightness of the outer (visible)
wall of the torus is only
$\sim 26,000$~K.
While this estimate is highly uncertain, the true
brightness temperature is unlikely to be more than
a factor of 2 -- 3 higher and is far less than
coronal temperatures.
We have, therefore, interpreted the torus as the 
optically-thick base of the disk wind, not as the optical
counterpart of the X-ray ADC.

We modeled the large hump in the orbital 
light curve of 4U~1822--371 with
a combination of heated face of the secondary
star and a variable-height, optically-thick wall
located at the edge of the disk. 
All previous models for the orbital light curve 
have needed a similar wall, but we now
suggest that the wall is actually
slightly closer to the neutron star than the 
edge of the disk and is part of the disk wind, not
the disk rim.
The wall is a cooler, darker layer at the base of
the disk wind just above the disk.

Our results thus yield a more complex view of the 
structure of the disk in 4U~1822--371.
The previous X-ray observations have shown that there 
is a vertically-extended, optically-thick ADC
stretching roughly half way out to the edge of the disk.
Our UV/optical observations show that most of the
outer disk emits a cooler disk wind.
The wind is is optically thick close to the accretion
disk but
becomes optically thin at a height of $\sim 0.5r_{\rm disk}$
above the disk.
Close to the disk the wind is yet cooler and darker,
forming a relatively dark wall around disk.
The disk itself appears to be nearly entirely buried 
within and
obscured by these vertically-extended structures.

\acknowledgments We thank Raymundo Baptista and Coel
Hellier for their valuable insights into this work.



\clearpage

\clearpage

\begin{table}
\centering
\begin{minipage}{4.5in}
\caption[Eclipse times and O-C] {Times of $T_{\rm min}$ and O-C values as
calculated using a quadratic least squares fit to the ephemeris.
\label{time}}
\begin{tabular}{lllr}    \hline \hline
Eclipse Time (HJD) & Error in $T_{\rm min}$ & Ref. & $O-C$ \\ \hline
2444044.8450 & 0.0060 & 1 & -0.0207 \\ 2444090.1140 & 0.0080 & 1 &
-0.0128 \\ 2444101.0280 & 0.0080 & 1 & -0.0079 \\ 2444105.6650 &
0.0060 & 1 & -0.0131 \\ 2444106.5970 & 0.0060 & 1 & -0.0095 \\
2444137.9350 & 0.0100 & 1 & -0.0062 \\ 2444411.1320 & 0.0080 & 2 &
-0.0007 \\ 2444412.0580 & 0.0080 & 2 & -0.0031 \\ 2444664.8120 &
0.0080 & 3 & -0.0152 \\ 2444783.8940 & 0.0060 & 2 & -0.0048 \\
2445579.5650 & 0.0030 & 4 & -0.0019 \\ 2445580.7250 & 0.0030 & 4 &
-0.0025 \\ 2445615.3115 & 0.0020 & 4 & -0.0002 \\ 2445937.7110 &
0.0021 & 4 & 0.0004 \\ 2446234.5787 & 0.0018 & 4 & 0.0010 \\
2447296.4798 & 0.0010 & 4 & 0.0040 \\ 2447379.3410 & 0.0010 & 4 &
0.0022 \\ 2447999.7674 & 0.0039 & 5 & 0.0011 \\ 2449163.0960 & 0.0028
& 6 & -0.0020 \\ 2449164.0237 & 0.0028 & 6 & -0.0027 \\ 2449164.2542 &
0.0028 & 6 & -0.0043 \\ 2449165.1852 & 0.0028 & 6 & -0.0018 \\
2449166.1190 & 0.0028 & 6 & 0.0036 \\ 2450250.5308 & 0.0008 & 7 &
-0.0003 \\ 2450250.7626 & 0.0008 & 7 & -0.0006 \\ 2450252.6196 &
0.0008 & 7 & -0.0004 \\ 2451264.8446 & 0.0035 & 5 & -0.0060 \\
2451373.7094 & 0.0008 & 7 & -0.0007 \\ 2451756.4577 & 0.0012 & 7 &
-0.0016 \\ 2451782.4542 & 0.0012 & 7 & -0.0015 \\ 2452089.7692 &
0.0008 & 7 & -0.0001 \\ 2453618.6767 & 0.0050 & 8 & -0.0023 \\
2453828.9720 & 0.0010 & 8 & 0.0010 \\ 2453830.8299 & 0.0010 & 8 &
0.0020 \\ 2453932.7201 & 0.0040 & 8 & -0.0042 \\
\end{tabular}
\tablenotetext{}{(1) \citet{mas80}, (2) \citet{mas82c}, (3)
  \citet{cow82}, (4) \citet{hel89}, \\ (5) \citet{cow03}, (6)
  \citet{har97}, (7) \citet{bap02}, (8) This paper}
\end{minipage}
\end{table}

\begin{table}
\center
\caption{Parameters of 4U 1822--371 as determined by model light curves.
\label{model}}
\begin{tabular}{lc} \hline \hline
  Parameter             & Value \\ 
  \hline 
  Period (days)         & 0.232109 \\ 
  Inclination (degrees) & $81.17\pm0.14$    \\ 
  $M_{NS}$ (\Msun)  (fixed)    & 1.35     \\ 
  $q=M_2/M_{NS}$        & 0.2971   \\ 
  $H_{\rm pow}$             & 1.1      \\ 
  $r_{\rm disk}$\tablenotemark{1}             & $0.4\pm0.01$  \\  
  \hline 
  UV Torus & \\
  Location, $a_{0}$\tablenotemark{1} & 0.165\\ 
  Torus Width\tablenotemark{1}   & 0.283\\ 
  Torus Edge Height\tablenotemark{1}  & 0.201\\ 
\\
  $V$ Torus & \\
  Location, $a_{0}$\tablenotemark{1}& 0.245\\ 
  Torus Width\tablenotemark{1}& 0.282\\ 
  Torus Edge Height\tablenotemark{1}  & 0.205\\ 
\\  
$J$ Torus & \\
  Location, $a_{0}$\tablenotemark{1}& 0.243\\ 
  Torus Width\tablenotemark{1}& 0.288\\ 
  Torus Edge Height\tablenotemark{1}  & 0.204\\ 
  \tablenotetext{1}{In units of the orbital separation (see Section 5.1.2).}
\end{tabular}
\end{table}

\end{document}